\documentclass[manuscript,screen,hidelinks]{acmart}
\AtBeginDocument{%
  }

\acmISBN{978-1-4503-XXXX-X/2018/06}

\usepackage{amsmath,amssymb}
\usepackage{graphicx}
\usepackage{textcomp}
\usepackage{xcolor}
\usepackage{booktabs}
\usepackage[table]{xcolor}
\usepackage[utf8]{inputenc}
\usepackage{xurl}
\usepackage{etoolbox}
\usepackage{pifont}
\usepackage{dashrule}
\usepackage[most]{tcolorbox}
\usepackage{makecell}
\usepackage{alphalph}
\usepackage{array}
\usepackage{geometry}
\usepackage{caption}
\usepackage{subcaption}
\usepackage{amsmath}
\usepackage{colortbl}
\usepackage{adjustbox}
\usepackage{longtable}
\usepackage{tabularx}
\usepackage{algorithm}
\usepackage{algpseudocode}
\definecolor{darkgreen}{rgb}{0.0, 0.5, 0.0}
\definecolor{codegreen}{rgb}{0,0.6,0}
\definecolor{codegray}{rgb}{0.5,0.5,0.5}
\definecolor{codepurple}{rgb}{0.58,0,0.82}
\definecolor{backcolour}{rgb}{0.95,0.95,0.92}
\definecolor{lightblue}{RGB}{235,245,255}
\definecolor{rowgray}{gray}{0.92}
\usepackage{amsmath}
\usepackage{natbib}
\usepackage{enumitem}
\usepackage{siunitx}
\usepackage{hyperref}
\usepackage{tikz}
\usepackage{multirow}
\usepackage{xspace}
\usepackage{rotating}
\usepackage{listings}
\usepackage{pgfplots}
\pgfplotsset{compat=1.18}
\usetikzlibrary{
  positioning,
  arrows.meta,
  shapes,
  shapes.geometric,
  shapes.misc,
  shapes.symbols,
  fit,
  calc
}
\lstdefinestyle{mystyle}{
    backgroundcolor=\color{backcolour},   
    commentstyle=\color{codegreen},
    keywordstyle=\color{magenta},
    numberstyle=\tiny\color{codegray},
    stringstyle=\color{codepurple},
    basicstyle=\footnotesize\ttfamily,
    breakatwhitespace=false,         
    breaklines=true,                 
    captionpos=b,                    
    keepspaces=true,                 
    numbers=left,                    
    numbersep=5pt,                  
    showspaces=false,                
    showstringspaces=false,
    showtabs=false,                  
    tabsize=2
}

\newcommand{\colorentry}[1]{%
    \ifdim #1 pt > 0 pt
        \textcolor{darkgreen}{#1}%
    \else
        \ifdim #1 pt < 0 pt
            \textcolor{red}{#1}%
        \else
            \textcolor{black}{#1}%
        \fi
    \fi
}

\definecolor{lightgreen}{RGB}{220, 255, 220}
\definecolor{lightred}{RGB}{255, 220, 220}
\definecolor{lightgray}{RGB}{240,240,240}
\definecolor{skyblue}{RGB}{135,206,235}
\definecolor{tealcolor}{RGB}{0,128,128}
\pgfplotsset{compat=1.18}

\newcommand{\good}[1]{\cellcolor{lightgreen}{#1}}

\newcommand{\colorsq}[1]{\colorbox{#1}{\phantom{X}}}

\begin{document}

\title{Exploring Statistical Change Point Detection Techniques for Performance
Anomaly Detection at Mozilla}

\newcommand{\diego}[1]{\textcolor{blue}{~Diego says:~#1}}
\newcommand{\bilel}[1]{\textcolor{red}{Bilel:#1}}
\newcommand{\tocite}[1]{\textcolor{red}{[#1]}}
\newcommand{\todo}[1]{\textcolor{red}{[#1]}}
\newcommand{\alexandersays}[1]{\textcolor{magenta}{~Alexander says:~#1}}
\newcommand{\philipsays}[1]{\textcolor{green}{~Philip says:~#1}}

\newcommand{\summarybox}[2]{%
  \begin{tcolorbox}[
    colback=lightblue,
    colframe=black,
    boxrule=0.4pt,
    arc=3pt,
    left=6pt, right=6pt, top=6pt, bottom=6pt,
  ]
  \textbf{#1}\\[3pt]
  #2
  \end{tcolorbox}
}

\newcommand{\practitionerquote}[2]{%
  \begin{tcolorbox}[
    blanker,
    borderline west={2pt}{0pt}{blue!60},
    left=10pt, right=0pt, top=1pt, bottom=0pt,
  ]
  \textit{``#1''} \hfill\textcolor{blue!70!black}{\small --- #2}
  \end{tcolorbox}
}

\newcommand{\rqone}{To what extent does \currentmozillamethod produce false alerts and miss real ones?~\xspace}

\newcommand{\rqtwo}{How effective are CPD methods in detecting performance changes at Mozilla?}

\newcommand{\rqthree}{Which CPD methods and ensemble techniques do Mozilla practitioners consider suitable for deployment in \textit{Perfherder}?}

\author{Mohamed Bilel Besbes}
\email{m_besbes@live.concordia.ca}
\affiliation{%
  \institution{REALISE Lab @ Concordia University}
  \city{Montréal}
  \state{Québec}
  \country{Canada}
}

\author{Gregory Mierzwinski}
\email{sparky@mozilla.com}
\affiliation{%
  \institution{Mozilla}
  \city{Potton}
  \state{Québec}
  \country{Canada}
}

\author{Suhaib Mujahid}
\email{smujahid@mozilla.com}
\affiliation{%
  \institution{Mozilla}
  \city{Montréal}
  \state{Québec}
  \country{Canada}
}

\author{Philipp Leitner}
\email{philipp.leitner@chalmers.se}
\affiliation{%
  \institution{Chalmers University of Technology and the University of Gothenburg}
  \city{Göteborg}
  \country{Sweden}
}

\author{Alexander Serebrenik}
\email{a.serebrenik@tue.nl}
\affiliation{%
  \institution{Eindhoven University of Technology}
  \city{Eindhoven}
  \state{North Brabant}
  \country{The Netherlands}
}

\author{Dave Hunt}
\email{dhunt@mozilla.com}
\affiliation{%
  \institution{Mozilla}
  \city{Kent}
  \state{England}
  \country{United Kingdom}
}

\author{Diego Elias Costa}
\email{diego.costa@concordia.ca}
\affiliation{%
  \institution{REALISE Lab @ Concordia University}
  \city{Montréal}
  \state{Québec}
  \country{Canada}
}

\renewcommand{\shortauthors}{Besbes, Mierzwinski, Mujahid, Leitner, Serebrenik, Hunt, and Costa}
\renewcommand{\arraystretch}{1.1}
\begin{abstract}

Software performance regressions can have significant business consequences, making automated detection a critical component of modern continuous integration pipelines. At Mozilla, performance anomaly detection is handled by \textit{Perfherder}, Mozilla's performance engineering management system that relies on a Student's T-test-based approach to flag regressions across hundreds of daily code changes. However, our preliminary analysis of one year of Mozilla performance data reveals that 12.5\% of generated alert groups are false positives, while approximately 6.8\% of them contain regressions missed by the automated system.
This paper presents an empirical study evaluating 25 change-point detection (CPD) methods and 15 ensemble approaches as alternatives to Mozilla's current method. We construct a ground-truth dataset of 174 performance time series manually annotated by eleven Mozilla performance engineers, representing one of the first practitioner-annotated CPD benchmarks for performance engineering. 
Our results show that while offline and hybrid CPD methods improve recall over Mozilla's method, they do so at a high cost to precision. Ensemble voting strategies alleviate this trade-off and offer more consistent performance, resulting in 11\% improvement in the F1-score. We validate the experimental results through a practitioner survey and report on lessons learned from integrating the best methods into Mozilla's performance engineering system. 

\end{abstract}


\received{20 February 2007}
\received[revised]{12 March 2009}
\received[accepted]{5 June 2009}

\maketitle

\newcommand{\currentmozillamethod}{Mozilla's method\xspace}

\section{Introduction}
\label{sec:introduction}

Slow and poorly optimized software can reduce customer satisfaction, increase operational costs, and drive users away from using a software product~\cite{perfenmgnfr}. It has been reported that a 1-second delay can cost \textit{Amazon} 1.6 Billion USD in sales~\cite{amazon16bloss}. In fact, a report by \textit{Google}, \textit{Deloitte}, and \textit{55} showcase that a 0.1-second site responsiveness improvement results in 8\% more user traffic and 10\% more revenue~\cite{google2020milliseconds}. To ensure software performance complies with user expectations, large software companies employ a rigorous process for testing performance~\cite{continuous_validation}, often under the umbrella of \textit{performance engineering}~\cite{stress_spike_testing}.

In large software development companies such as Mozilla, developers commit hundreds of code changes every day. Measuring and monitoring the impact of such changes on the software performance is inherently challenging. Designing reliable performance tests is difficult, as tests must be both representative of real-world workloads and reproducible across executions~\cite{Costa:JMHBenchmarks, continuous_validation}.
Software performance is known to be influenced by external factors (e.g., hardware non-determinism), which introduce noise that is hard to control~\cite{envnoiserelatednondeterminism, perf_change_detected_in_history, leitnernoisehandling}. 
As a result, distinguishing genuine performance regressions from measurement noise is a major challenge in performance engineering, particularly since performance changes often manifest subtly at runtime and may go unnoticed during code review~\cite{perf_test_is_noisy, code_changes_intro_perf_regressions, beller2023learninglearnpredictperformance}.

Therefore, reliable performance testing requires human orchestration and dedicated tooling to manage the performance engineering workflow~\cite{liu2023humaninthelooponlinejustintimesoftware, perfabnormalities, markusse2022using, muehlbauer2019accurate, lee2012}. 
A notable example is \textit{MongoDB}, which uses the E-Divisive change point detection algorithm to automatically flag potential regressions across hundreds of benchmarks~\cite{mongodb_cpd}. 
Mozilla follows a similar model through~\textit{Perfherder}~\cite{perfherder}, its own performance engineering management system. 
It offers a wide selection of performance tests across multiple environmental configurations. 
It analyzes performance measurement history to detect potential regressions using a statistical change point detection approach~\cite{student1908}. Detected regressions are surfaced as \textit{performance alerts}, which are manually triaged by performance engineering experts known as \textit{Performance Sheriffs} and dispatched to the responsible development teams for remediation.

Despite this structured process, change point detection is inherently imperfect as no method can fully eliminate false alerts or missed ones~\cite{cpd_survey}. Prior work on statistical change point detection for performance anomaly detection in industrial settings has reported the existence of both failure modes. \citet{mongodb_cpd} report that MongoDB's original detection system produced false positive rates as high as 99\%, while \citet{bipec} similarly document the co-existence of false and missed alerts in SAP HANA's regression detection pipeline. In large-scale continuous integration environments like Mozilla's, where hundreds of revisions are tested daily across a broad set of performance tests, even a modest proportion of such errors can translate to meaningful overhead for Performance Sheriffs and potential quality risk for end users. This motivates a systematic investigation into the limitations of Mozilla's current method and whether alternative statistical methods could offer improvements.

Our study answers the following research questions:
\begin{itemize}
    \item \textbf{RQ1:} \rqone
    \item \textbf{RQ2:} \rqtwo
    \item \textbf{RQ3:} \rqthree
\end{itemize}

To answer these questions, we structure this study as a design science one~\cite{designscience, Engstr_m_2020}, framing it as a problem-solution inquiry~\cite{problem_solution_pair_example_study_1, problem_solution_pair_example_study_2}. We first characterize the limitations of \currentmozillamethod, then propose and evaluate alternative solutions. We begin the characterization of \currentmozillamethod's limitations by quantifying the scale of false and missed alerts through a preliminary analysis of one year of Mozilla performance data. To conduct the evaluation, we build upon \textit{TCPDBench}~\cite{TCPDBenchAlanTuringInstitute}, an established CPD benchmarking toolset, which we extend with 15 previously unrepresented methods to cover online and hybrid families alongside the existing offline ones. We further construct a ground-truth dataset of 174 performance time series, annotated manually by eleven Mozilla performance engineers, to benchmark CPD methods against \currentmozillamethod in a realistic industrial setting. Finally, since strong benchmark performance alone is insufficient to justify deployment in an industrial setting, we complement our quantitative evaluation with a workshop and survey involving twelve Mozilla performance engineers. This allows us to validate whether the methods that perform best on our benchmark are also considered suitable for deployment by the practitioners who would use them, and whether the assumptions underlying our evaluation reflect the realities of Mozilla's performance testing workflow.
The main contributions of this study are as follows:

\begin{itemize}
    \item \textbf{A characterization of Mozilla's performance testing workflow}. Detailed accounts of performance engineering workflows from large-scale industrial systems are rare in the literature, making this characterization a valuable reference for researchers designing and evaluating CPD methods in realistic continuous delivery contexts, as well as for practitioners designing and improving performance engineering systems.
    \item \textbf{A practitioner-annotated ground-truth dataset of 174 performance time series}, labelled by eleven Mozilla engineers. 
    This is a large-scale CPD benchmark grounded in expert knowledge from a real-world continuous delivery environment. 
    The dataset and replication package are publicly available on Zenodo\footnote{https://doi.org/10.5281/zenodo.20382322} and can be used to foster further research in CPD methods.

    \item \textbf{An extension of the \textit{TCPDBench} evaluation toolset}~\cite{TCPDBenchAlanTuringInstitute} with 15 previously unrepresented CPD methods, including nine online methods, six hybrid methods, and a replication of \currentmozillamethod, enabling a broader assessment of the CPD landscape beyond offline methods alone.

    \item \textbf{A large-scale evaluation of CPD methods}, including 25 individual CPD methods and 15 ensemble voting strategies.
    Our results show that a selection of offline and hybrid methods consistently improves recall over the baseline, but at the cost of precision. 
    Ensemble methods, however, show promise, yielding better overall performance and lower false-alert rates.
    
\end{itemize}

The rest of this paper is structured as follows.
We ground our study in the Design Science paradigm (Section~\ref{sec:study_design}).
Section~\ref{sec:rqone} then investigates Mozilla's performance testing workflow and quantifies the scale of false and missing alerts.
In Section~\ref{sec:methodology}, we detail our experimental setup: the practitioner-annotated ground-truth dataset, CPD method selection, and ensemble voting strategy design.
Our evaluation results across offline, online, hybrid, and ensemble approaches are reported in Section~\ref{sec:results}.
Section~\ref{sec:industrial_validation} presents the practitioner survey, which validates our evaluation assumptions and proposed solutions with Mozilla engineers.
We discuss implications for researchers and practitioners, including a per-characteristic hyperparameter tuning analysis, in Section~\ref{sec:discussion}.
Threats to validity are addressed in Section~\ref{sec:threats_to_validity}, followed by related work in Section~\ref{sec:related_works}.
Section~\ref{sec:conclusion} concludes and outlines directions for future research.

\section{Study Design}
\label{sec:study_design}

\begin{figure}
    \centering
    \includegraphics[width=0.8\columnwidth]{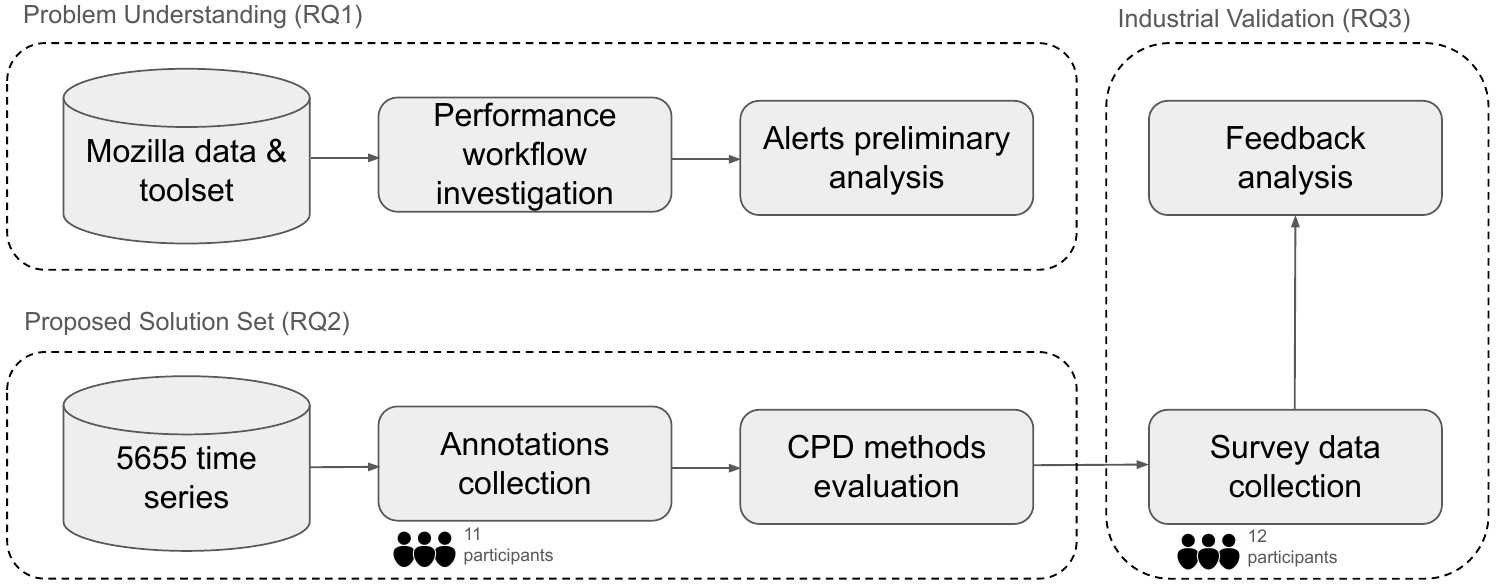}
    \caption{Study Design Overview}
    \label{fig:study_design_overview}
\end{figure}

We use the \textit{Design Science} paradigm to describe our research study. The design science paradigm focuses on the creation and rigorous evaluation of artifacts such as models, methods, and systems to solve identified problems and contribute to knowledge~\cite{designscience,Engstr_m_2020}. Our study falls into the \textit{Problem-solution} parity category, as we focus on understanding the problem instance (inaccuracies of the current method), and evaluate and propose a set of potential solutions (CPD methods).  
Figure \ref{fig:study_design_overview} provides a high-level idea of the research process, and in the remainder of the section, we describe the core building blocks of our study.  
\\

\noindent
\textbf{Problem Understanding.}
We investigate Mozilla’s source code and toolset, workflow documentation, and the dataset of Mozilla performance measurements~\cite {Besbes_2025} to develop an understanding of the Mozilla anomaly detection workflow (Section \ref{sub:performance_workflow_investigation}).
By analyzing these resources, we consolidate Mozilla’s performance engineering process and we evaluate the scale of the identified issues (Section \ref{sub:contribution_area_to_workflow}). 
We find that at least 12\% of \textit{alert summaries}, collections of alerts, are raised falsely. We also observe that Performance Engineers at Mozilla create alerts on their own in Mozilla's alert-tracking system, indicating performance changes that are large enough for engineers to manually intervene, rather than being detected by \currentmozillamethod. 
\\

\noindent
\textbf{Proposed Solution Set.}
We investigate academic literature to identify alternatives to \currentmozillamethod, as well as different strategies to leverage these alternatives mentioned in Section \ref{subsec:sota_methods}.
In order to evaluate the effectiveness of these methods, we conduct a large-scale annotation collection process from Mozilla engineers described in Section \ref{subsec:data_annotation_process}. We select a sample of 400 samples from the dataset containing 5655 time series as shown in Figure \ref{fig:study_design_overview}, and we deploy these time series on a platform so they could be annotated by Performance Engineers. Once the data collection period concludes, we obtain 174 annotated time series by eleven different participants. We integrate those annotations along with the implementation of the selected CPD methods into a tool called \textit{TCPDBench}. 
\\

\noindent
\textbf{Industrial Validation}
To complement our quantitative evaluation, we conduct a practitioner validation study with Mozilla performance engineers.

The validation pursues two complementary goals: validating the assumptions underlying our evaluation framework, and assessing the performance of the proposed solutions on \textit{Perfherder}.

We describe the validation design and procedure, followed by the survey instrument in Section~\ref{subsec:survey_instrument_design}, the survey demographics in Section~\ref{subsec:survey_demographics}, and the results in Sections for the research assumptions validation and the assessment of the proposed solutions' performance ~\ref{subsec:validating_research_assumptions} and~\ref{subsec:validating_cpd_results} respectively.

\section{\rqone}
\label{sec:rqone}

This section dives into the process adopted by Mozilla to identify performance anomalies during software development. 
We describe Mozilla's performance testing workflow in Section~\ref{sub:performance_workflow_investigation} and dive into a preliminary analysis of the issues of the current alert system in Section~\ref{sub:contribution_area_to_workflow}.

\subsection{The Mozilla's Performance Testing Workflow}
\label{sub:performance_workflow_investigation}

The goal of Mozilla's Performance Testing workflow is to capture and remediate performance regressions as part of the continuous integration and delivery process (CI/CD).
Mozilla's performance engineering workflow is orchestrated via a Mozilla-specific system called \textit{Perfherder}~\cite{perfherder}, the performance alerts management platform used at Mozilla. 
To mitigate performance test costs, performance tests are run at fixed time intervals, varying per project.   
For example, in the \textit{Autoland} project~\cite{autoland}, performance tests are run every four hours, always testing the latest working revision of the software project. 
The performance measurements are organized into time series and analyzed to identify performance anomalies. 
Any performance anomaly detected prompts the creation of a \textit{performance alert}. 
If multiple performance alerts are created for the same revision, they are grouped into \textit{performance alert summaries}, to facilitate analysis by the performance sheriffs.

\begin{figure}
    \centering
    \begin{subfigure}[t]{0.4\textwidth}
        \centering
        \includegraphics[height=3cm]{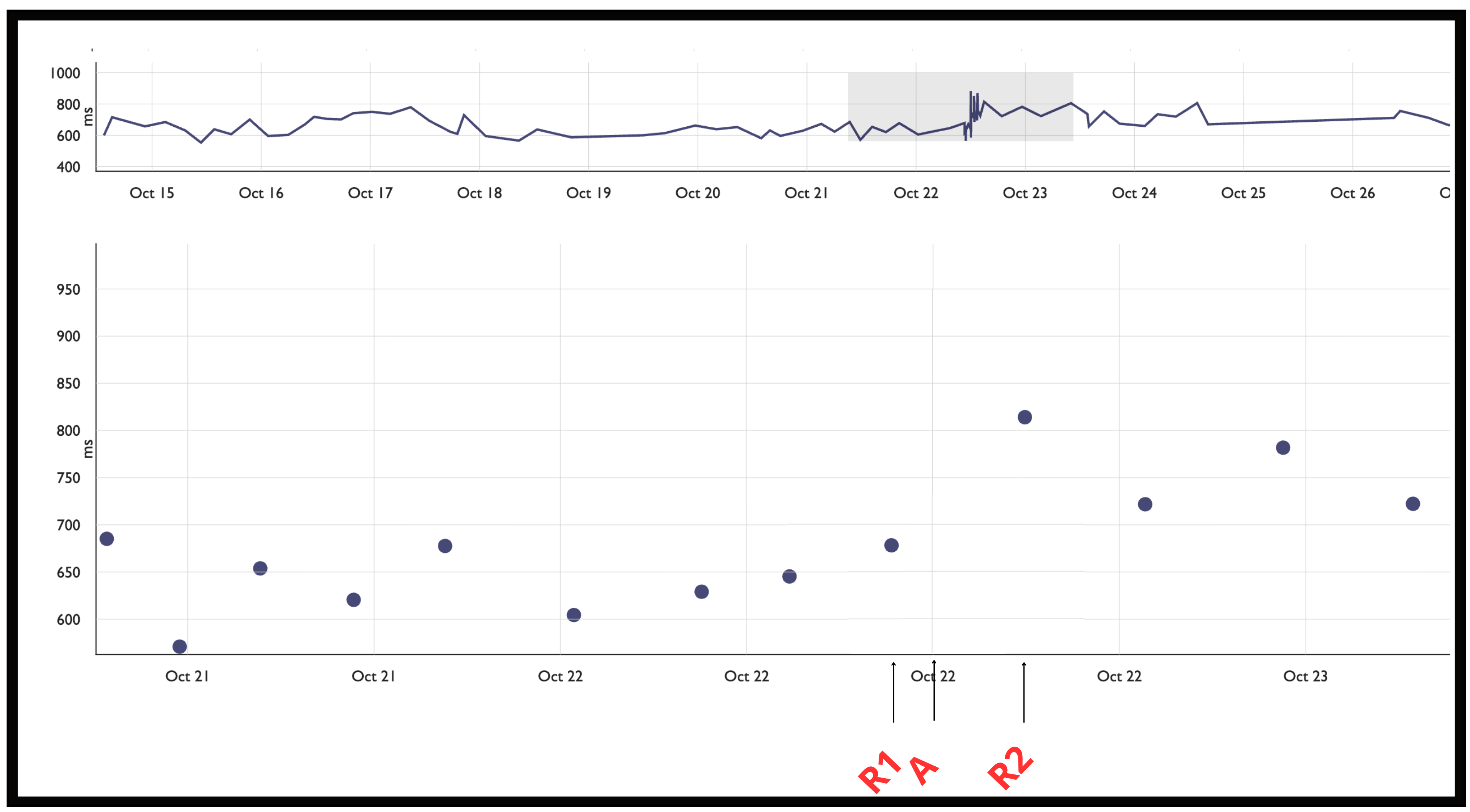}
        \caption{Regular intermittent performance evaluation}
        \label{fig:regular_run}
    \end{subfigure}%
    \begin{subfigure}[t]{0.14\textwidth}
        \centering
        \includegraphics[height=3cm]{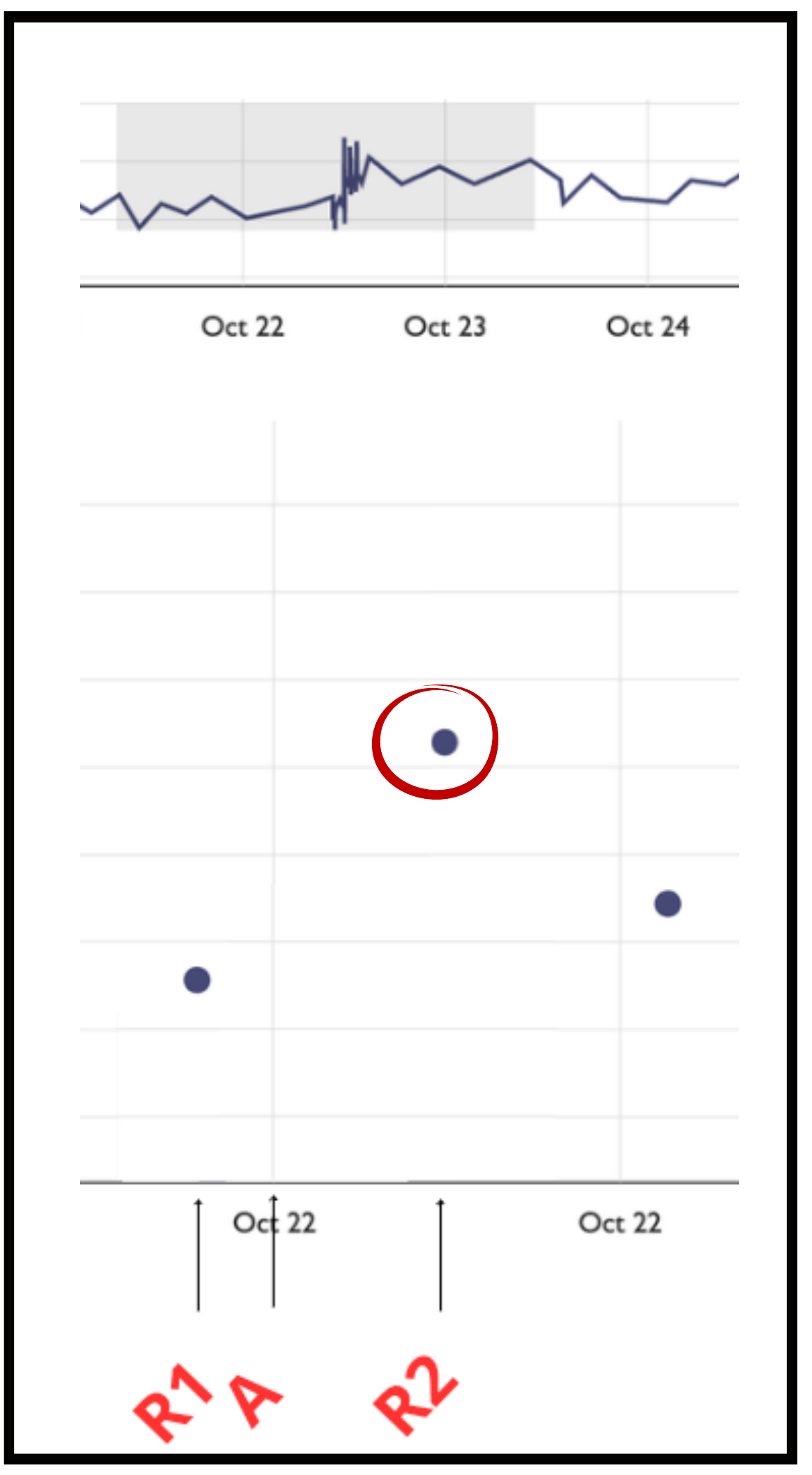}
        \caption{Anomaly \\ detection}
        \label{fig:anomaly_detection}
    \end{subfigure}%
    \begin{subfigure}[t]{0.14\textwidth}
        \centering
        \includegraphics[height=3cm]{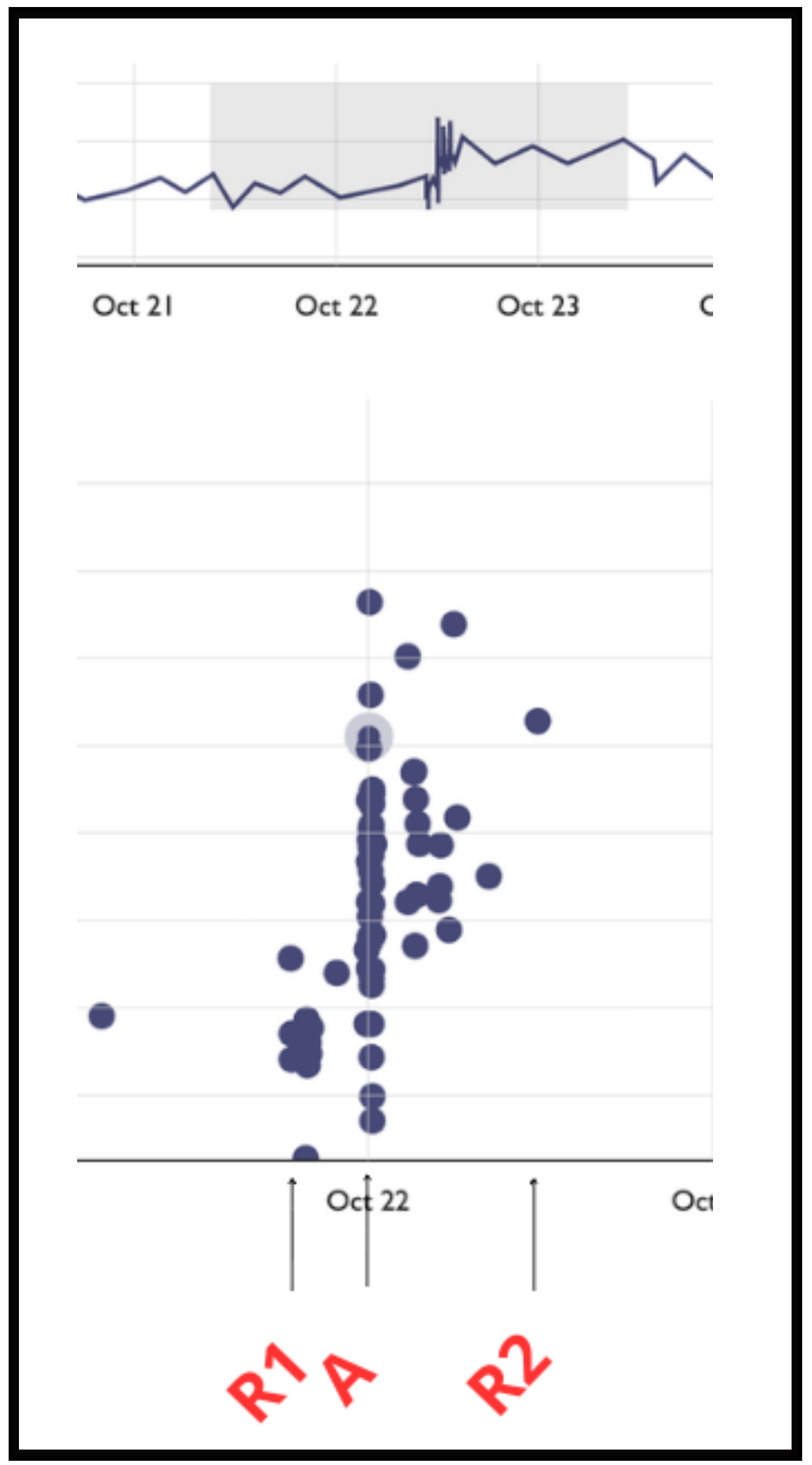}
        \caption{Backfilling \\ \& Retriggering}
        \label{fig:backfilling_retriggering}
    \end{subfigure}%
    \begin{subfigure}[t]{0.3\textwidth}
        \centering
        \includegraphics[height=3cm]{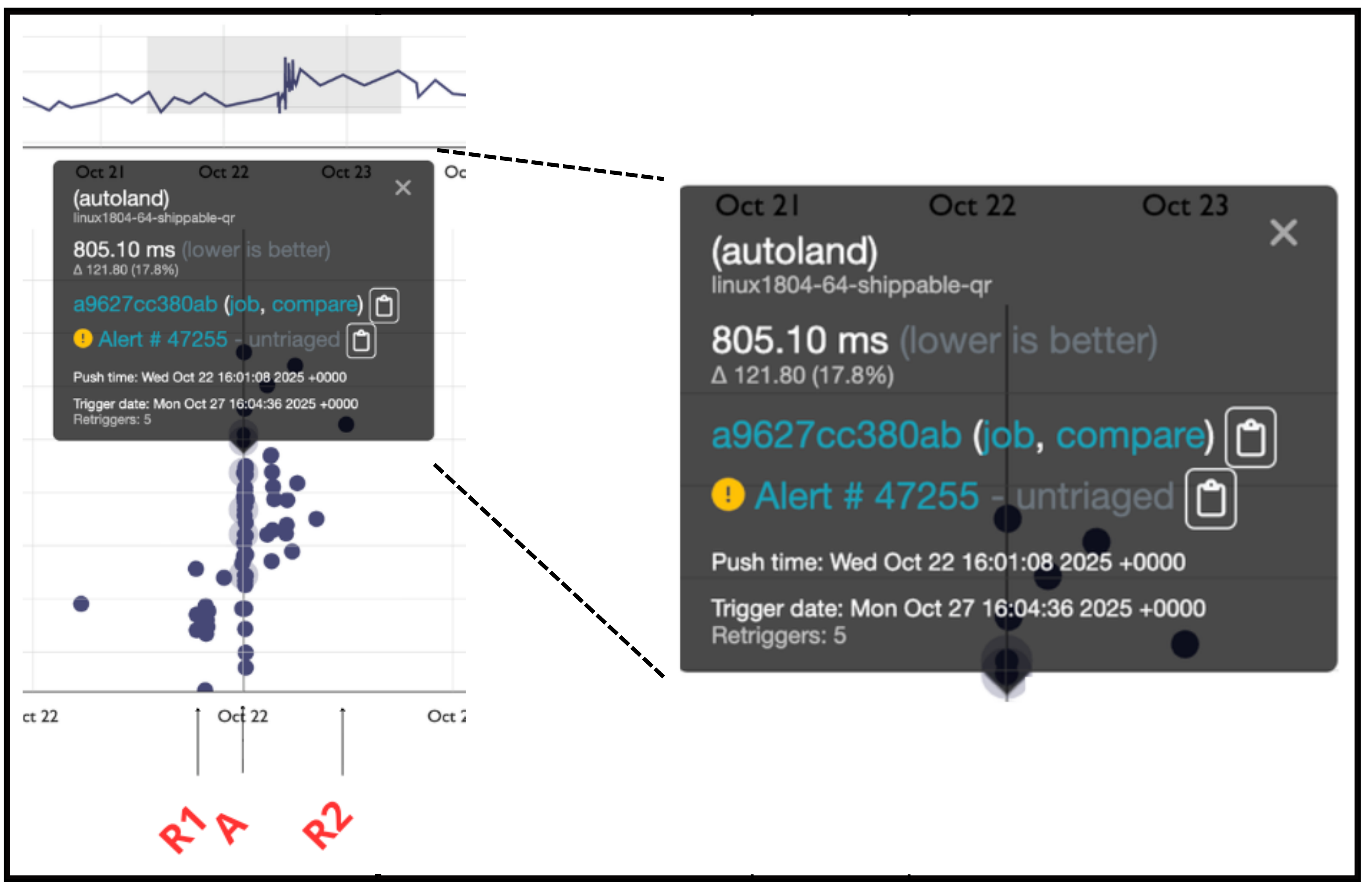}
        \caption{Fine-granular anomaly localization}
        \label{fig:backfill_retrigger_alert_creation}
    \end{subfigure}

    \caption{Alert creation example}
    \label{fig:alert_creation_example}
\end{figure}

To elaborate on the performance anomaly detection, we present an illustrative example in Figure~\ref{fig:alert_creation_example}.
Specifically, Figure~\ref{fig:regular_run} shows historical performance measurements of a target system, spanning three days of measurements. 
A noticeable performance change is observed between revisions \textit{R1} and \textit{R2} through running a Student-$t$-test based analysis on performance measurements, which is highlighted in Figure~\ref{fig:anomaly_detection}.
Given that multiple revisions may exist between \textit{R1} and \textit{R2}, \textit{Perfherder} starts a process called \textit{backfilling} \& \textit{retriggering}. This approach of isolating performance regressions is well-established in the performance testing community, having been adopted in performance engineering industrial settings over a decade ago~\cite{agg_perf_measurements}.

These processes ensure that any revisions pushed between \textit{R1} and \textit{R2} are performance-tested and measured \textbf{multiple times} as performance testing requires multiple testing runs on the same setup in order to reduce the effect of noise coming from hardware nondeterminism~\cite{code_changes_intro_perf_regressions, taming_performance_variability}. Once the new runs are done, the Student-$t$-test-based analysis is run again so the culprit revision, which is \textit{A} in our case, is identified.
Figure~\ref{fig:backfilling_retriggering}, which now shows a higher resolution of measurements between \textit{R1} and \textit{R2}. 
Once the culprit revision is identified, an alert associated with it is created as shown in Figure \ref{fig:backfill_retrigger_alert_creation}.

\begin{figure}
\centering
\resizebox{\columnwidth}{!}{%
\begin{tikzpicture}[
    node distance=2.2cm,
    every node/.style={font=\normalsize, align=center},
    process/.style={rectangle, draw, rounded corners=2pt, minimum width=3cm, minimum height=1.1cm, thick},
    decision/.style={diamond, draw, aspect=2, inner sep=2pt, thick},
    startstop/.style={circle, draw, minimum size=0.8cm, thick},
    arrow/.style={-Stealth, thick}
]
\usetikzlibrary{positioning}

\node[startstop] (start) {\tikz \fill (0,0) circle (6pt);};
\node[below=1pt of start] {\small revision data\\[-2pt]\small \textasciitilde timeseries};

\node[process, right=of start] (filter) {Timeseries\\filtering};
\node[process, right=of filter] (tvalue) {T-value\\calculation};
\node[decision, right=of tvalue] (tcheck) {T-value $>$ 7?};

\node[process, below=1.8cm of tcheck] (compute) {Computing\\change magnitude};
\node[decision, left=of compute] (magcheck) {mag. $>$ 2\%?};
\node[process, left=of magcheck] (alert) {Alert\\creation};

\draw[arrow] (start) -- (filter);
\draw[arrow] (filter) -- (tvalue);
\draw[arrow] (tvalue) -- (tcheck);

\draw[arrow] (tcheck.east) -- ++(1.0,0) node[right, font=\normalsize] {Not an alert};
\node[above right=2pt and 6pt of tcheck.east, font=\small\bfseries] {No};

\draw[arrow] (tcheck.south) -- node[midway, right, font=\small\bfseries] {Yes} (compute.north);

\draw[arrow] (compute) -- (magcheck);
\draw[arrow] (magcheck) -- node[midway, above, font=\small\bfseries] {Yes} (alert);

\draw[arrow] (magcheck.south) -- ++(0,-0.8) node[below, font=\normalsize] {Not an alert};
\node[below right=2pt and 6pt of magcheck.south, font=\small\bfseries] {No};

\end{tikzpicture}%
}
\caption{Simplified workflow of Mozilla’s alert creation process}
\label{fig:cpd_&_alertrep}
\end{figure}
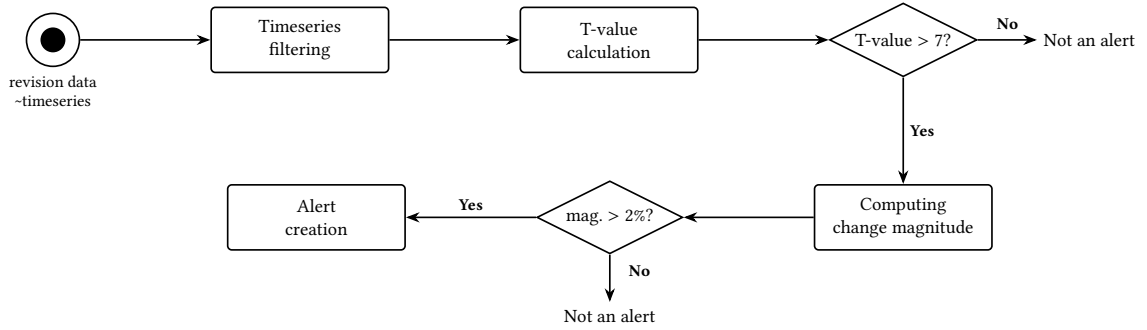

\currentmozillamethod of change point detection follows the pipeline described in Figure~\ref{fig:cpd_&_alertrep}.
The method is comprised of four steps:

    \textbf{Step 1. Select target measurements}: Each performance measurement is associated with a project revision, and is chronologically sorted based on revision dates.
    To investigate a performance anomaly related to \textit{R2}, the method selects a set of measurements from revisions pushed before \textit{R2} (\textit{pre}) and a set of measurements of revisions pushed after \textit{R2}, including \textit{R2} itself (\textit{post}). 
    It is customary to use 12 to 24 measurements for the \textit{pre} set and 12 measurements for the \textit{post} set.

    \textbf{Step 2. Compare \textit{pre} and \textit{post} sets using Student T test metrics}:
    The method then compares both \textit{pre} and \textit{post} sets using a Student $t$-test~\cite{student1908}. 
    Instead of relying on the usual p-values, the method instead reports on the computed T-value, as a threshold of interest. 
    If the T-value is strictly above the critical threshold of 7, the revision is marked as a \textit{suspected} anomaly. Otherwise, the given revision is discarded as a suspected performance anomaly, and the system moves on to test other revisions. In our example, the change introduced by \textit{A} has a T-value equal to 8.237, exceeding the change detection threshold.
    
    \textbf{Step 3. Compare change magnitude}: 
    The previous step's \textit{suspected} anomaly incurs further investigation because the performance measurements before it drastically differ in scale compared to it and ones subsequent to it. 
    Therefore, the magnitude of change between the two sets of measurements defined earlier is computed and compared to a critical value of 2\%. In case it surpasses that threshold an anomaly is confirmed at \textit{R2} such as the example in Figure \ref{fig:anomaly_detection}. The practice of double-step change point verification has also been reported in SAP HANA's regression detection use case~\cite{bipec}.

    \textbf{Step 4. Manual Triage.} 
    Alert summaries are manually investigated by Performance Sheriffs, who conduct a root cause analysis to determine whether the alert corresponds to a genuine performance regression. When a regression is confirmed, the Sheriffs file a bug report and assign it to the appropriate development team. They subsequently monitor the issue until it is properly addressed.

In summary, the process of detecting and precisely locating performance anomalies requires intermittent performance evaluation periodically, which creates the measurements used to evaluate performance changes, prompting the detection of a performance anomaly, as shown in Figure \ref{fig:anomaly_detection}. Depending on the magnitude of the performance change, it could be further analyzed to assign it to the exact revision that introduced the performance change and create an alert for it, as shown in Figure \ref{fig:backfill_retrigger_alert_creation}. Subsequently, we refer to the described CPD method as \textit{Mozilla's method} or simply as the \textit{current method}.

\summarybox{How does Mozilla identify performance regressions?}{Mozilla's performance testing workflow relies on \textit{Perfherder} to periodically run performance tests and detect anomalies via a Student's $t$-test-based change point detection pipeline. When a potential regression is detected between two tested revisions, backfills are triggered to identify the culprit revision and reduce measurement noise. Confirmed anomalies are grouped into alert summaries and triaged manually by Performance Sheriffs.}

\subsection{A Preliminary Analysis on Missing and False Alerting}
\label{sub:contribution_area_to_workflow}

To assess the frequency with which alerts are falsely generated or valid anomalies are missed, we conduct a preliminary analysis of one year of performance measurements at Mozilla. We collect a dataset using the Mozilla \textit{Perfherder} API spanning May 2023 to May 2024. The dataset contains 17,989 performance alerts generated from performance tests across multiple software platforms and systems. These alerts correspond to 5,655 performance time series across 186 different test suites and are grouped into 3,912 unique alert summaries.

\noindent
\paragraph{False alerting analysis}
False alerting is manifested as alert summaries that appear on \textit{Perfherder}, which do not correspond to actual performance anomalies. They could be caused by infrastructure issues or by high noise levels compared to the size of the change that was detected.
Such summaries have to be triaged by performance sheriffs, and a high rate of false alerts cost time, effort, and may lead to practitioners distrusting the alert system. Bug \#1977385\footnote{https://bugzilla.mozilla.org/show\_bug.cgi?id=1977385} on BugZilla is associated with a false alert summary, which required approximately four business days of investigation by an engineer, ultimately revealing that the flagged issue is not an issue to begin with.

In order to identify the scale of false alerting, we begin by classifying each alert summary into one of three categories based on the status assigned to it by Performance Sheriffs:

\begin{itemize} 
    \item \textbf{True Alert summaries:} Represent validated performance anomalies or regressions. This category includes alerts with statuses such as \textit{fixed, improvement, reassigned, backedout, downstream,} or \textit{wontfix}, where \textit{wontfix} alert summaries could correspond to valid regressions spotted in performance measurements but not reflected in the product itself due to configuration issues. 
    
    \item \textbf{False Alert summaries:} Do not represent real performance issues and often result from noise or irrelevant environmental factors. We identify these alerts by their \textit{invalid} status.

    \item \textbf{Uncertain Alert summaries:} Summaries where the validity remains undetermined and requires further review. These are labeled with statuses such as \textit{investigating} or \textit{untriaged} 
\end{itemize}

\begin{table}[t]
\centering
\caption{Distribution of Alerting Classifications}
\label{tab:alert_summary_classification}
\begin{tabular}{l r r}
\hline
\textbf{Classification} & \textbf{Count} & \textbf{Percentage} \\
\hline
True Alert summaries    & 2203 & 56.3\% \\
False Alert summaries     & 488  & 12.5\% \\
Uncertain Alert summaries & 1221 & 31.2\% \\
Summaries with Missed Alerts & 266 & 6.8\% \\
\midrule
Total Unique Summaries     & 3912 & 100\%  \\
\bottomrule
\end{tabular}
\end{table}

As shown in Table \ref{tab:alert_summary_classification}, our preliminary analysis of these alert summaries indicates that 56.3\% are classified as true alerts, 12.5\% as false, and 31.2\% as uncertain.
This means that at least one eighth of generated performance alert summaries are false.

\paragraph{Missing alerting analysis}
Performance alerts can also be manually created by Performance Sheriffs when the automated tooling misses an issue. A manually created alert signals that the underlying performance change was not captured by \textit{Perfherder} and that the regression was significant enough to be noticed and flagged by a human engineer. Notably, a small observed rate of such human-detected misses may plausibly understate the true prevalence of undetected issues, particularly when practitioners place high trust in the automated detection system~\cite{fn_motivation_1, trust_in_seq_non_param_tests, trust_in_auto_study}. Our analysis reveals that 6.8\% of alert summaries contain at least one manually created alert, suggesting that missed detections, while infrequent, remain a non-trivial concern.

\paragraph{Conclusion}
Overall, our analysis reveals that
\textit{Perfherder} exhibits notable limitations in both false positive and missed detection rates. 
One possible factor contributing to these issues is the statistical unfitness of the Student's $t$-test for the use case in which it is currently employed. The test assumes data normality and equal variance between the samples being compared. However, these assumptions do not necessarily hold for software performance measurements~\cite{taming_performance_variability}. Such measurements are often noisy and may follow non-normal distributions.
Therefore, we hypothesize that, although the chosen method is generally sound, it may not always be fully aligned with the characteristics of the underlying data in this context.

\summarybox{How often does the~\currentmozillamethod miss or issue false alerts?}{Our analysis of 3,912 alert summaries spanning one year of Mozilla performance data reveals that at least 12.5\% are false alerts, while 6.8\% contain manually created alerts indicating missed detections by the automated system. A further 31.2\% remain unresolved. This motivates the investigation of more robust detection methods.}

\section{\textbf{Experimental Setup}}
\label{sec:methodology}

\begin{figure}
    \centering
    \includegraphics[width=0.9\columnwidth]{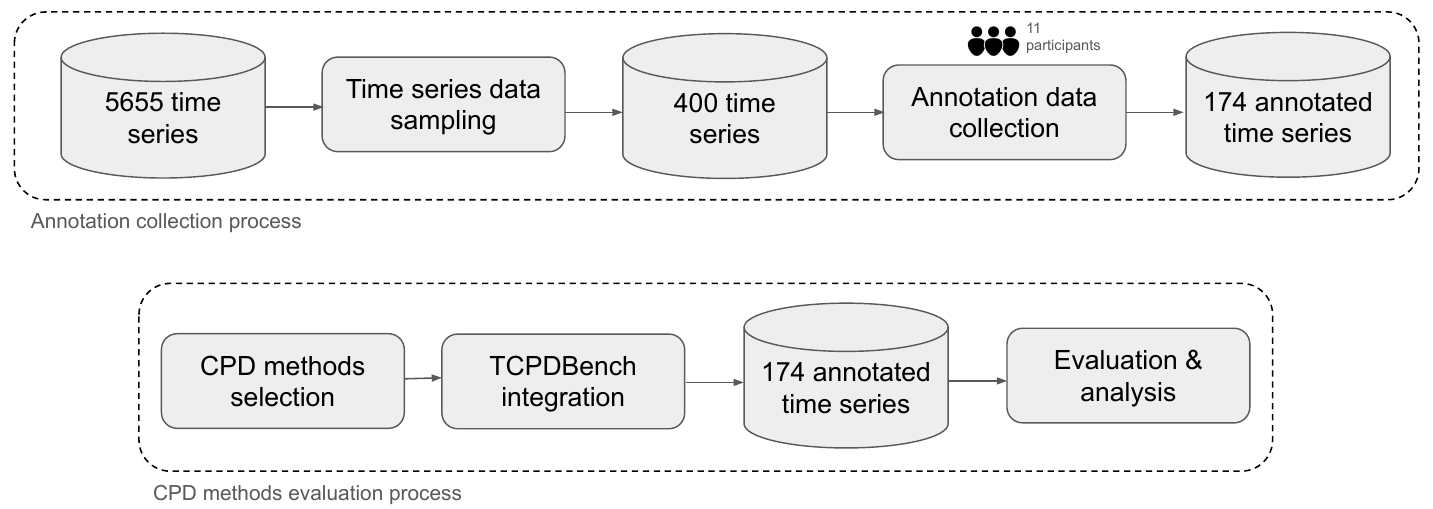}
    \caption{Experimental methodology overview}
    \label{fig:rq2_methodology}
\end{figure}

After exploring the limitations of \currentmozillamethod, we conduct an experimental study to explore alternative change-point detection (CPD) methods. 
The goal is to determine whether CPD methods can provide greater reliability than \currentmozillamethod by minimizing false and/or missing alerts.
Figure~\ref{fig:rq2_methodology} illustrates the experimental methodology used to collect and annotate the data, select the alternative CPD methods, and execute the experiments.

\subsection{Dataset Sample Selection}
\label{subsec:dataset_sample_selection}

To evaluate the CPD techniques in our study, we require a reliable ground-truth dataset. Mozilla’s existing alert summaries cannot be used directly for this purpose for two reasons. First, using alerts produced by \currentmozillamethod as a ground-truth risks confirmation bias, potentially reinforcing its assumptions and limitations. Second, the data associated with \textit{backfilling} and \textit{retriggering}, two processes that launch multiple tests upon finding a potential alert to determine the exact culprit revision, makes it impossible to distinguish original measurements from those produced by the latter processes. Since no such gold-standard labeling exists in Mozilla’s performance pipeline, we need to construct one. 
This requires collecting practitioner annotations on selected time series, providing a reliable reference against which different change point detection methods can be compared.
Guided by the Mozilla performance engineering team, from a published dataset\cite{Besbes_2025}, we select time series from the two most reliable testing subsets, which are called \textit{Speedometer3} and \textit{TP6}, respectively.
By selecting time series associated with these subsets among the entire published dataset of 5655 time series, we obtain 2851 time series. From those, we randomly sample 400 for the annotation process to maintain a 95\% confidence level and a 5\% margin of error.

In the performance time series dataset, some revisions are associated with multiple performance measurements (up to 20 measurements for a single revision in certain cases) while others have only one measurement per revision.
To simplify the annotation process for human annotators, we represent each revision with exactly one measurement in the annotation platform. When a revision has multiple associated measurements, we aggregate them by averaging\cite{agg_perf_measurements, mongodb_virtuous_cycle}, so that each revision is displayed as a single representative value. The annotation process is described in detail in Section~\ref{subsec:data_annotation_process}.

\subsection{Annotation Platform}
\label{subsec:annotation_platform_selection}

\begin{figure}
    \centering
    \includegraphics[width=0.5\columnwidth]{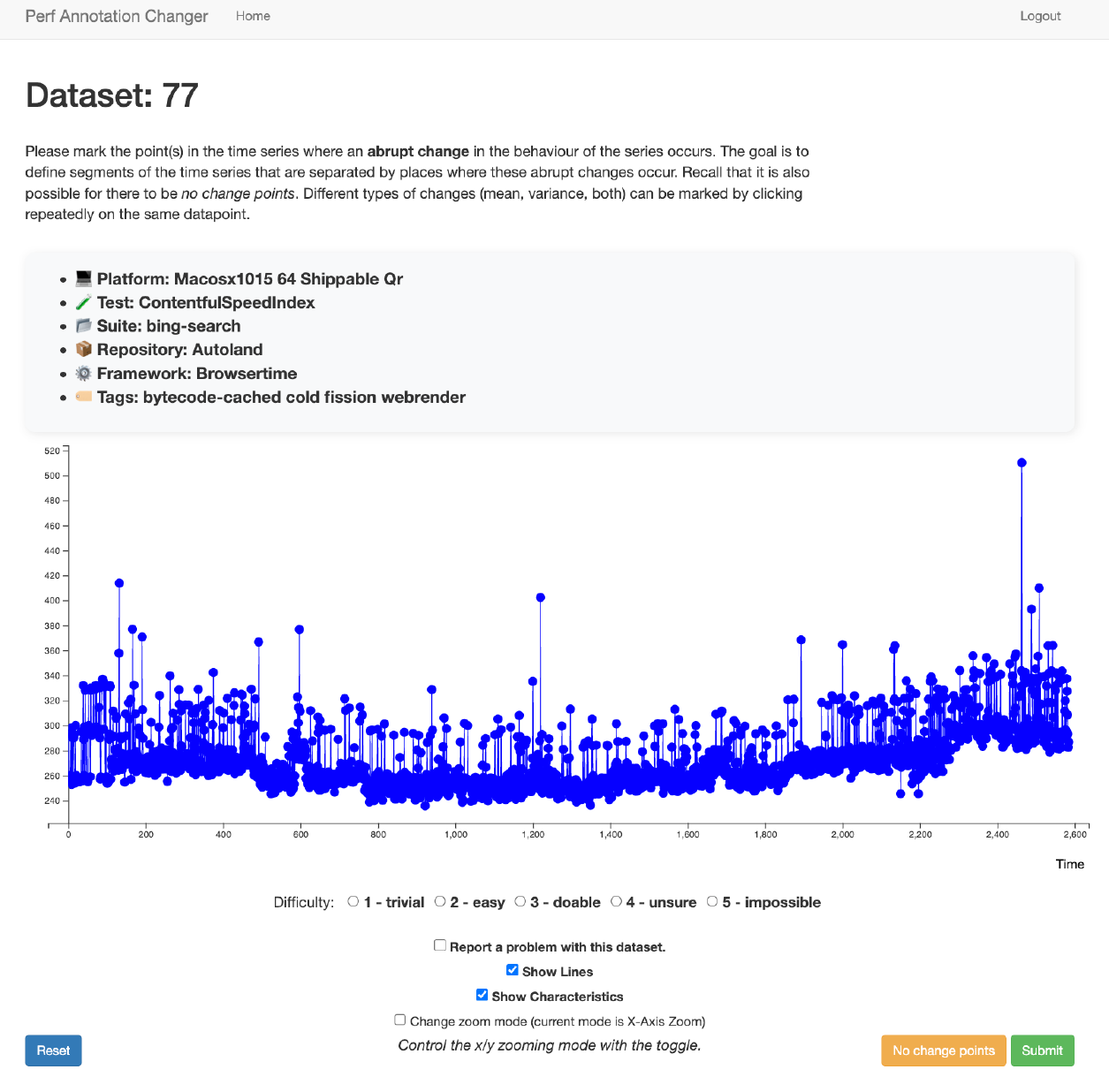}
    \caption{Annotation user interface example}
    \label{fig:annotation_ui_example}
\end{figure}

To conduct the time series annotation process, we need a tool that allows participants to interactively explore the time series and annotate potential performance changes. 
For that aim, we deploy an open-source application called  \textit{AnnotationChange}\footnote{\url{https://github.com/alan-turing-institute/AnnotateChange}}.
This tool is a web platform created by ~\citet{TCPDBenchAlanTuringInstitute} and is used in previous studies to collect time-series annotations from participants~\cite{atashgahi2023memoryfreeonlinechangepointdetection, 101093imaiaiiaaf002}.
\textit{AnnotationChange} also has a simpler but familiar interface to Mozilla's time series dashboard, which facilitates annotator onboarding by using a familiar setup.
Figure~\ref{fig:annotation_ui_example} shows an example of the annotation interface. Out of the box, the tool offers the following features:

\begin{itemize}
    \item \textbf{Customized tutorials:} We design a tutorial that includes an initial onboarding process.
    
    \item \textbf{Zooming/sliding options:} When annotating time series with carrying lengths, annotators must have the option to zoom in and out and to move in four directions within the time series plot.

    \item \textbf{Automated management of annotation tasks.} 
    \textit{AnnotationChange} automatically manages the assignment of annotation tasks among annotators. It prioritizes time series that have already received at least one annotation and continues assigning them to additional annotators until the specified maximum number of annotators per time series is reached. The system also prevents the same annotator from labeling the same time series more than once.
\end{itemize}

Based on the feedback from a pilot study with two researchers and two practitioners, we add more detailed tutorial content and performance test specifications, and add capabilities to zoom on the y-axis to the tool.

\subsection{Data Annotation Process}
\label{subsec:data_annotation_process}

We include the performance time series in the annotation platform without any prior alert-related labels. 
The goal of this process is to annotate these time series by identifying performance changes that should be notified to performance Sheriffs.
The resulting annotations will then be used as ground truth for evaluating the CPD methods. 
To increase the reliability of the annotated ground-truth data, we configure the annotation platform to assign five annotators for each performance time series.

Participants are recruited through an internal email sent to members of Mozilla’s performance engineering teams. The invitation includes a video explaining the study's motivation, the goals of the annotation process, and brief instructions for using the annotation platform. Upon registration, participants are required to complete a mandatory tutorial before annotating any of the time series used in this study.

A total of eleven Mozilla practitioners volunteered to join the data annotation process. Table \ref{tab:demographics} presents the demographic information of the participants, including their roles, academic backgrounds, and experience with software engineering and Mozilla’s performance engineering workflow. 
Participants have extensive experience with performance engineering, with an average of 13 years of software engineering experience, five years of experience with Mozilla’s performance engineering workflow, and four years of experience using \textit{Perfherder}.

\begin{table}
\centering
\caption{Demographic information about the participants, including their roles, overall software engineering experience, and familiarity with the Mozilla performance workflow.}
\begin{tabular}{l l r r r}
\toprule
& & \multicolumn{3}{c}{\textbf{Years of Experience}} \\
\textbf{Role} & \textbf{Academic level} & \textbf{Engineering} & \textbf{Mozilla Performance} & \textbf{Perfherder} \\
\midrule
Perf Test Engineer & Masters & 10 years & 7 years & 7 years \\

Perf Sheriff \& Perf Testing & Masters & 13 years & 13 years & 5 years \\

SE Intern & Bachelor & 2 years & 0.5 years & 0.5 years \\

Engineering Manager & College & 25 years & 5 years & 5 years \\

Perf Engineer & Bachelor & 24 years & 7 years & 7 years \\

Perf Engineer & Bachelor & 24 years & 3 years & 3 years \\

Perf Sheriff & High School & 6 years & 5 years & 5 years \\

Perf Sheriff \& Perf Test Engineer & Masters & 8 years & 2 years & 2 years \\

Perf Sheriff & Bachelor & 6 years & 5 years & 5 years \\

Perf Sheriff \& Perf Test Engineer & Bachelor & 14 years & 4 years & 4 years \\

\midrule
\textbf{Average} & \textbf{---} & \textbf{13.2 years} & \textbf{5.1 years} & \textbf{4.3 years} \\
\bottomrule
\end{tabular}
\label{tab:demographics}
\end{table}

The dataset deployed in the platform initially contained 400 instances of performance time series.
During the one-month annotation period, we observe that completing annotations for all 400 time series with multiple annotators would be infeasible given the time-intensive nature of the task and the restricted pool of eligible engineers.
We therefore choose to concentrate annotations on a smaller subset, prioritizing the original goal of five annotators, rather than having fewer annotators on a large sample of time series.
Consequently, we concentrate the annotation on 174 time series, allowing annotators to focus on a smaller set while increasing the number of annotations per time series.   

\subsection{Selecting Change Point Detection methods}
\label{subsec:sota_methods}

Using the annotated dataset, we aim to evaluate the performance of various CPD methods. 
We select the CPD methods by surveying the academic literature and identifying approaches commonly used for detecting changes in time series data~\cite{Ross01042012, RAAB2020340, welch, cvm}. 
In addition, we include \currentmozillamethod. 
We base our evaluation on \textit{TCPDBench}, the benchmarking framework designed to evaluate and compare CPD techniques on time series datasets.
For the methods that are not available on \textit{TCPDBench}, we implement and integrate them into it~\cite{TCPDBench_simon_replication_package}.
Integrating these methods into \textit{TCPDBench} ensures that they can be evaluated under a consistent experimental setup. Furthermore, the extended version of \textit{TCPDBench} developed in this study is fully reproducible and publicly available, allowing future research to easily reuse the implementations and replicate our evaluation~\cite{replicationPackage}.

We choose a broad range of CPD methods in an attempt to explore various setups and select the best for Mozilla's performance measurements. 
We detail the selection process per CPD method category. Table \ref{tab:cpd_methods_hyperparam} summarizes all methods chosen in this study.
\\

\textbf{Offline methods.}
Offline CPD methods analyze the entire time series at once to identify points at which statistical properties change~\cite{cpd_selective_reviw}. These methods are particularly useful when all data has already been collected, and changes need to be detected retrospectively with high accuracy~\cite{offline_online}. 
We begin with the set of offline CPD methods proposed in the study introducing \textit{TCPDBench}~\cite{TCPDBenchAlanTuringInstitute}. Following a preliminary analysis, we exclude the following methods: \textit{Energy Change Point}~\cite{james2013ecprpackagenonparametric}, and \textit{Prophet}~\cite{forecasting_at_scale} due to implementation issues and lack of compatibility with our experimental setup. 
The final set of methods considered in our evaluation is summarized in Table~\ref{tab:cpd_methods_hyperparam}.
In addition, we incorporate the \textit{E-Divisive} method, which has been used in production settings such as at MongoDB~\cite{mongodb_cpd}.

\textbf{Online methods}
Online methods operate in a streaming setting, where observations are processed sequentially as they arrive. At each time step, the method evaluates the incoming data point in the context of previously observed data and, in some cases, reassesses earlier points within a retained history window to determine whether a change has occurred, without access to future observations~\cite{cpd_survey, bocpd}. The objective is to detect changes as quickly as possible after they occur, such as in cases of anomaly detection in vital infrastructure metrics~\cite{usertrafficanomalydetection, Chen_2022} where quick detection is needed, at the cost of being less precise than offline methods~\cite{offline_online, li2019reviewchangepointdetectionmodels}

\begin{table*}
\centering
\caption{Summary of Change Detection Methods}
\label{tab:cpd_methods_hyperparam}
\footnotesize

\begin{tabular}{l l l l r}
\toprule
\textbf{Category} & \textbf{Abbrev.} & \textbf{Full Name} & \textbf{Implementation} & \textbf{\# H-param. Combos} \\
\midrule

\rowcolor{rowgray}
Offline & AMOC     & At Most One Change~\cite{amoc}                              & \textit{changepoint} R~\cite{changepoint}         & 126 \\
        & BinSeg   & Binary Segmentation~\cite{binseg}                          & \textit{changepoint} R~\cite{changepoint}         & 252 \\
\rowcolor{rowgray}
        & CPNP     & Nonparametric Change Point Detection~\cite{fearnhead2017changepointdetectionpresenceoutliers}  & \textit{changepoint} R~\cite{changepoint}         & 28  \\
        & KCPA     & Kernel Change-Point Analysis~\cite{kcpa}                   & \textit{ecp} R~\cite{james2013ecprpackagenonparametric}                 & 18  \\
\rowcolor{rowgray}
        & MongoDB  & E-Divisive~\cite{mongodb_cpd}                              & \textit{spa} Python~\cite{e_divisive}             & 12  \\
        & PELT     & Pruned Exact Linear Time~\cite{pelt}                       & \textit{changepoint} R~\cite{changepoint}         & 126 \\
\rowcolor{rowgray}
        & RFPOP    & Robust Functional Pruning Optimal Partitioning~\cite{fearnhead2017changepointdetectionpresenceoutliers}& \textit{robustFPOP} R~\cite{robust-fpop}          & 4   \\
        & SegNeigh & Segment Neighborhoods~\cite{segneigh}                      & \textit{changepoint} R~\cite{changepoint}         & 252 \\
\rowcolor{rowgray}
        & WBS      & Wild Binary Segmentation~\cite{wbs}                        & \textit{wbs} R~\cite{wbs_package}                 & 12  \\
        & Zero     & No Change Points                                           & Custom                                            & --  \\

\midrule

\rowcolor{rowgray}
Online  & CUSUM        & Cumulative Sum Control Chart~\cite{cusum}              & Custom                                            & 750 \\
        & EWMA         & Exponentially Weighted Moving Average~\cite{ewma}      & Custom                                            & 108 \\
\rowcolor{rowgray}
        & Page Hinkley & Page Hinkley Test~\cite{cusum,Sebastio2017SupportingTP}            & \textit{river} Python~\cite{montiel2021river}     & 432 \\
        & ADWIN *        & Adaptive Windowing~\cite{adwin}                        & \textit{river} Python~\cite{montiel2021river}     & 180 \\
\rowcolor{rowgray}
        & BOCPD *        & Bayesian Online Changepoint Detection~\cite{bocpd}     & \textit{ocp} R~\cite{bocpdrepack}                 & 81  \\
        & Shewhart     & Online Shewhart Control Chart~\cite{shewhart}          & Custom                                            & 27  \\
\rowcolor{rowgray}
        & KSWIN *        & Kolmogorov-Smirnov Windowing~\cite{RAAB2020340}        & \textit{river} Python~\cite{montiel2021river}     & 36  \\
        & CVMWIN *       & Cramér-von Mises Windowing~\cite{Ross01042012}               & \textit{alibi-detect} Python~\cite{alibidetect}   & 48  \\
\rowcolor{rowgray}
        & SPRT         & Sequential Probability Ratio Test~\cite{sprtogpubliction}          & Custom                                            & 27  \\
        & ODummy       & Dummy Drift Detector~\cite{odummy}                     & \textit{river} Python~\cite{montiel2021river}     & -- \\

\midrule

\rowcolor{rowgray}
Hybrid  & CVM    & Cramér-von Mises Test~\cite{cvm}       & \texttt{scipy.cramervonmises\_2samp}~\cite{cramervonmises_2samp} & 792 \\
        & WELCH  & Welch's $t$-test~\cite{welch}             & \texttt{scipy.ttest\_ind}~\cite{ttest_ind}               & 792 \\
\rowcolor{rowgray}
        & MWU    & Mann-Whitney U Test~\cite{mwu}          & \texttt{scipy.mannwhitneyu}~\cite{mannwhitneyu}          & 792 \\
        & LEVENE & Levene's Test~\cite{levene}             & \texttt{scipy.levene}~\cite{leveneimplementation}        & 792 \\
\rowcolor{rowgray}
        & KS     & Kolmogorov-Smirnov Test~\cite{ks}       & \texttt{scipy.ks\_2samp}~\cite{ks_2samp}                 & 792 \\

\bottomrule
\end{tabular}
\vspace{2pt}
\raggedright
\footnotesize{\textbf{*} Detection may be attributed to an earlier point within the retained observation window.}
\end{table*}

\begin{figure}
\centering
\resizebox{\columnwidth}{!}{%
\begin{tikzpicture}[
    node distance=2.2cm,
    every node/.style={font=\normalsize, align=center},
    process/.style={rectangle, draw, rounded corners=2pt, minimum width=3cm, minimum height=1.1cm, thick},
    decision/.style={diamond, draw, aspect=2, inner sep=2pt, thick},
    startstop/.style={circle, draw, minimum size=0.8cm, thick},
    arrow/.style={-Stealth, thick},
    legend/.style={rectangle, draw, rounded corners=2pt, dashed, inner sep=6pt,
                   font=\small, align=left, text=gray, draw=gray}
]
\usetikzlibrary{positioning}

\node[startstop] (start) {\tikz \fill (0,0) circle (6pt);};
\node[below=1pt of start] {\small revision data\\[-2pt]\small \textasciitilde timeseries};
\node[process, right=of start]  (filter)     {Timeseries\\filtering};
\node[process, right=of filter] (stattest)   {Statistical\\test};
\node[decision, right=of stattest] (sigcheck) {$p$-value $< \alpha$?};
\node[process,  below=1.8cm of sigcheck] (effectcomp)  {Magnitude of\\change check};
\node[decision, left=of effectcomp]      (effectcheck) {Mag. Change $> \tau$?};
\node[process,  left=of effectcheck]     (alert)        {Alert\\creation};

\node[legend, above=1.0cm of filter] (legwin) {%
    \textbf{Window parameters}\\
    min win $\in \{\textcolor{red}{12}, 18, 24\}$\\
    max win $\in \{\textcolor{red}{24}, 36, 48\}$\\
    fwd win $\in \{3, 6, \textcolor{red}{12}\}$};
\draw[dashed, gray, -Stealth] (legwin.south) -- (filter.north);

\node[legend, above=1.4cm of stattest] (leg1) {%
    \textbf{Statistical tests}\\
    MWU \textbullet{} CVM \textbullet{} Welch\\
    Levene \textbullet{} KS};
\draw[dashed, gray, -Stealth] (leg1.south) -- (stattest.north);

\node[legend, above=1.0cm of sigcheck] (legalpha) {%
    \textbf{Significance threshold}\\
    $\alpha \in \{0.005, 0.01, 0.02,$\\
    $0.035, \textcolor{red}{0.05}, 0.06,$\\
    $0.07, 0.08, 0.1\}$};
\draw[dashed, gray, -Stealth] (legalpha.south) -- (sigcheck.north);

\node[legend, above=0.4cm of effectcheck] (legtau) {%
    \textbf{Magnitude threshold}\\
    $\tau \in \{1, \textcolor{red}{2}, 3\}$};
\draw[dashed, gray, -Stealth] (legtau.south) -- (effectcheck.north);

\draw[arrow] (start)    -- (filter);
\draw[arrow] (filter)   -- (stattest);
\draw[arrow] (stattest) -- (sigcheck);
\draw[arrow] (sigcheck.east) -- ++(1.0,0)
    node[right, font=\normalsize] {Not an alert};
\node[above right=2pt and 6pt of sigcheck.east, font=\small\bfseries] {No};
\draw[arrow] (sigcheck.south) --
    node[midway, right, font=\small\bfseries] {Yes} (effectcomp.north);
\draw[arrow] (effectcomp)  -- (effectcheck);
\draw[arrow] (effectcheck) --
    node[midway, above, font=\small\bfseries] {Yes} (alert);
\draw[arrow] (effectcheck.south) -- ++(0,-0.8)
    node[below, font=\normalsize] {Not an alert};
\node[below right=2pt and 6pt of effectcheck.south, font=\small\bfseries] {No};

\end{tikzpicture}
}
\caption{Hybrid CPD methods pipeline breakdown. We highlight the varying statistical tests and hyperparameter search space. Default values are highlighted in red.}
\label{fig:hybrid_methods_breakdown}
\end{figure}
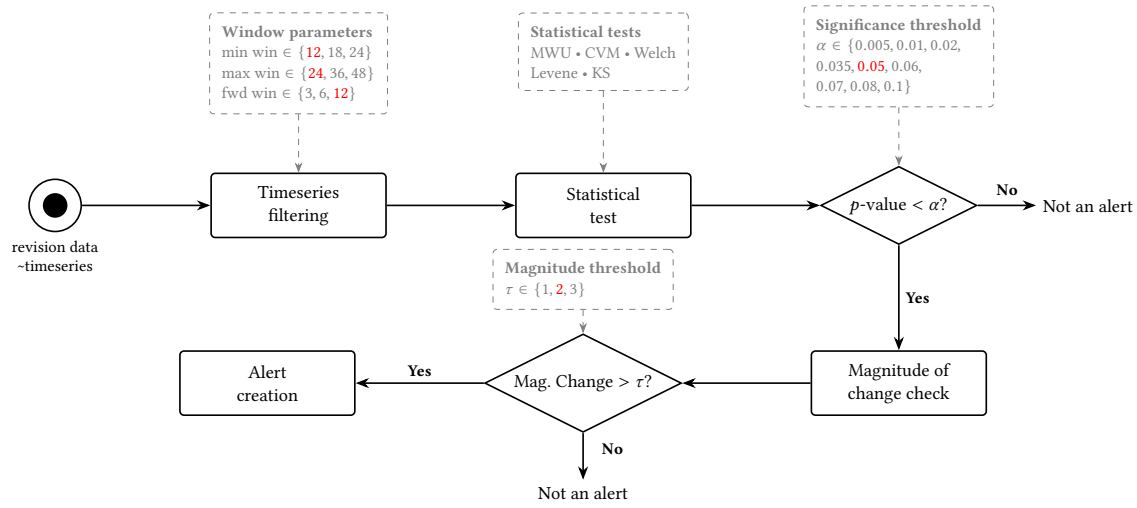

\textbf{Hybrid CPD Methods.}
In addition to the literature-selected methods, we propose a set of hybrid CPD methods inspired by \currentmozillamethod. 
Each hybrid method evaluates a candidate revision comparing two local windows of measurements, one preceding and one following the revision, using a statistical test method, which is itself a hybrid method.
As shown in Figure~\ref{fig:hybrid_methods_breakdown}, the statistical test and effect size measure are interchangeable components, controlled by parameters $\alpha$ and $\tau$ respectively.
The default window sizes and magnitude threshold match those of \currentmozillamethod, and the p-value is set to \textbf{0.05}, following common practice~\cite{Wasserstein02042016, demsar2006statistical, arcuri2011statistical}.   \\

\textbf{TCPDBench integration.}
For offline methods, we retain those from the original replication package.
We newly introduce nine online detection techniques, as well as the hybrid methods described in Section~\ref{subsec:sota_methods}, which share the same parameter space as \currentmozillamethod (critical value, window sizes).
We also replicate \currentmozillamethod~\cite{TCPDBench_simon_replication_package} and validate it against 30 Mozilla performance time series cleaned of backfilling runs, confirming that the results match \textit{Perfherder} exactly. We also introduce five hybrid detection techniques.
Finally, we configure TCPDBench to include all 174 annotated time series and their ground-truth change point labels.

\subsection{Evaluation Metrics}
\label{subsec:metrics}

Change point detection can be cast as a binary classification problem, where each time step $t \in [1, T]$ is classified as either a change point or a non-change point\cite{pelt, cpd_survey}. 
Because change points typically constitute a small fraction of all time steps, however, standard accuracy is heavily skewed toward the majority class and is therefore uninformative. 
We instead evaluate methods using \textbf{precision}, the proportion of predicted change points that correspond to true change points, and \textbf{recall}: the proportion of true change points that are successfully detected. 
We combine both into the \textbf{F\textsubscript{1}-score}, which weights precision and recall equally, reflecting our goal of minimizing both false positives and false negatives.

Change point detection inherently involves some degree of ambiguity in the exact location of a change: a detection that is off by a few time steps may still be practically meaningful. It is therefore common to define a margin of error $M$ around each true change point location, within which a predicted change point is counted as a true positive~\cite{pelt, cpd_selective_reviw}. We use $M = 5$ throughout following prior work\cite{TCPDBenchAlanTuringInstitute}.

Our setting introduces an additional challenge: each time series is annotated by multiple users, and annotators may disagree on the number or location of change points. Following the framework proposed by~\citet{Martin2004} and adopted by~\citet{TCPDBenchAlanTuringInstitute}, we adapt precision and recall to account for multiple ground-truth annotations. Let $T_k$ denote the set of change points annotated by user $k$, for $k = 1, \ldots, K$, let $X$ denote the set of predicted change points, and define
\[
    TP(A, B) = \bigl\{ \tau \in A \;\big|\; \exists\, x \in B 
    \text{ such that } |\tau - x| \leq M,\ 
    \text{with each } x \in B \text{ matched to at most one } \tau \bigr\}.
\]
Letting $T^* = \bigcup_{k=1}^{K} T_k$ be the union of all annotations, precision is the fraction of predictions matched by at least one annotator
\begin{equation}
    P = \frac{|TP(T^*, X)|}{|X|}
    \label{eq:precision}
\end{equation}
Recall is the average, across all annotators, of how many of their annotations were detected: 
\begin{equation}
    R = \frac{1}{K} \sum_{k=1}^{K} \frac{|TP(T_k, X)|}{|T_k|}
    \label{eq:recall}
\end{equation}

Averaging recall across annotators, rather than pooling all annotations, ensures that each annotator's perspective contributes equally~\cite{TCPDBenchAlanTuringInstitute}.
\begin{equation}
    F_1 = \frac{2PR}{P + R}
    \label{eq:f1_score}
\end{equation}

As commonly adopted in change point detection in time series~\cite{bipec, TCPDBenchAlanTuringInstitute}, we include $t = 1$ as a trivial change point in every $T_k$ and in $X$ to prevent division by zero when an annotator or method reports no change points,

\subsection{Experiment Setup}
\label{subsec:setup}

Each CPD method considered in this study exposes a set of hyper-parameters that control its behavior. As these parameters vary across methods (e.g., window sizes, thresholds, penalty terms), we perform hyperparameter tuning by defining, for each method, a corresponding search space of possible configurations. The size of these search spaces, expressed as the number of hyperparameter combinations explored per method, is reported in Table~\ref{tab:cpd_methods_hyperparam}. This tuning process allows us to systematically explore the parameter space, assess the sensitivity of each method to its configuration, and identify settings that yield optimal performance.
 
To evaluate the impact of hyperparameter tuning, we report results under three complementary evaluation setups, following prior work~\cite{TCPDBenchAlanTuringInstitute}:

\begin{itemize}
    \item \textbf{Default Configuration:} For a given method, we use its default hyperparameter configuration and compute the average metric values across all datasets.
    
    \item \textbf{Best Configuration:} For a given method, we perform hyperparameter tuning by evaluating all configurations in its search space and computing the average F1-score across all datasets for each configuration. We then select the configuration that achieves the highest average F1-score. The corresponding precision and recall values associated with this configuration are reported as \textit{Best Precision} and \textit{Best Recall}.
    
    \item \textbf{Upper-Bound Performance:} For a given method, we select, for each dataset independently, the hyperparameter configuration that yields the best performance. We then average these best-per-dataset metric values across all datasets for each metric separately. This setup represents an upper bound on performance when optimal hyper-parameters are chosen with full knowledge of the dataset.
\end{itemize}

\subsection{Ensemble of Methods}
\label{subsec:ensemble_of_methods_methodology}

\begin{figure}
    \centering
    \includegraphics[width=0.7\columnwidth]{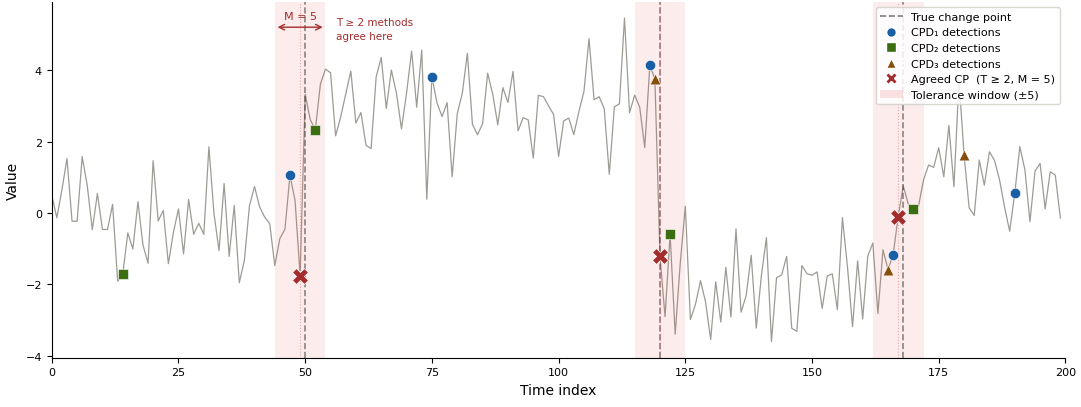}
    \caption{CPD agreement example: consensus threshold $C$ of 2, tolerance margin M of 5}
    \label{fig:agreement_example}
\end{figure}

We want to evaluate the effectiveness of combining multiple CPD methods using an ensemble voting strategy. 
The intuition is that performance changes reported by multiple distinct CPD methods are more likely to be true positives. 
Since different methods may not flag the same event at the exact same index, we employ a \textit{tolerance margin} $M$, which defines a window within which detections from different methods are treated as referring to the same event.
Then, we define a \textit{consensus threshold} $C$, which specifies the minimum number of methods that must agree on for a performance change to be issued.  
Figure~\ref{fig:agreement_example} illustrates this on a simulated time series with three true change points: three CPD methods each produce their own set of detections, and a collective agreement is declared only where at least $C=2$ methods detect a change within a tolerance window of $M=5$, with the resulting agreed change point assigned to the average index of the contributing detections.

\section{\textbf{\rqtwo}}
\label{sec:results}

This section presents the assessment of the data collected during the annotations process as well as the results of applying CPD methods to identify the annotated changes using offline, online, hybrid, and ensemble CPD methods. 
The results are compared against the performance obtained with \currentmozillamethod as well as dummy baselines specific to each CPD method family.

\subsection{Annotation Data Quality Assessment}
\label{subsec:annotation_data_quality_assessment}

Our data annotation experiment lasted for approximately one month. In this period, eleven Mozilla annotators participated in the experiment, annotating 174 timeseries. 
We present some general statistics in Table~\ref{tab:general_stat}.
Annotators took, on average, 5.5 minutes to finish a task, with half the tasks completed in around 1 minute. A task is the action of annotating one time series by one annotator.
On average, an annotator marks 6 to 7 change points per timeseries, though 61 tasks received no annotations whatsoever, meaning those timeseries were deemed free of any performance anomaly. 
Some annotators are notably more active than others, ranging from 1 to 2 change points per task on average to as many as 14, possibly reflecting differences in sensitivity to subtle changes. 
Annotators also assigned a difficulty level to each time series among five levels ranging from trivial to impossible: 69.7\% were considered at least doable, while in 30.3\% of cases, annotators were unsure or found it impossible to distinguish changes from noise. 

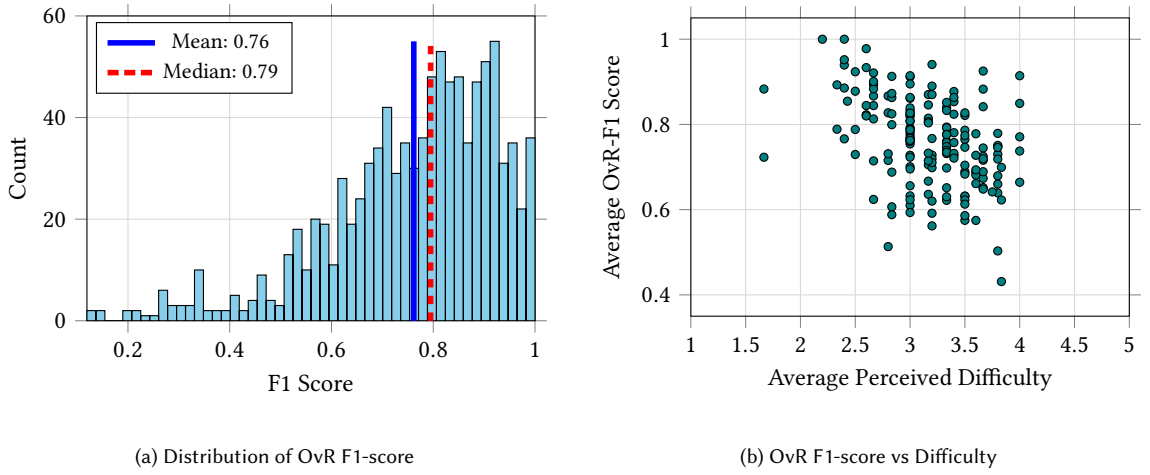
\begin{figure}
\centering
\begin{subfigure}[b]{0.48\textwidth}
\centering
\resizebox{\linewidth}{!}{\begin{tikzpicture}
  \begin{axis}[
    title={},
    xlabel={F1 Score},
    ylabel={Count},
    title style={font=\fontsize{11}{13}\selectfont},
    label style={font=\fontsize{9}{11}\selectfont},
    tick label style={font=\fontsize{9}{11}\selectfont},
    legend style={
      font=\fontsize{8}{10}\selectfont,
      at={(0.02,0.98)},
      anchor=north west,
    },
    xmin=0.11938139, xmax=1.00000000,
    ymin=0, ymax=60,
    bar width=0.01761237,
    width=\linewidth,
    height=0.75\linewidth,
    grid=both,
    grid style={line width=0.4pt, draw=gray!30},
    xtick align=outside,
    ytick align=outside,
  ]
 
  \addplot[
    ybar interval,
    fill={rgb,255:red,135;green,206;blue,235},
    draw=black,
    line width=0.4pt,
    forget plot,
  ] coordinates {
    (0.11938139,2)
    (0.13699377,2)
    (0.15460614,0)
    (0.17221851,0)
    (0.18983088,2)
    (0.20744325,2)
    (0.22505563,1)
    (0.24266800,1)
    (0.26028037,6)
    (0.27789274,3)
    (0.29550511,3)
    (0.31311749,3)
    (0.33072986,10)
    (0.34834223,2)
    (0.36595460,2)
    (0.38356698,2)
    (0.40117935,5)
    (0.41879172,2)
    (0.43640409,4)
    (0.45401646,9)
    (0.47162884,4)
    (0.48924121,3)
    (0.50685358,13)
    (0.52446595,18)
    (0.54207832,10)
    (0.55969070,20)
    (0.57730307,19)
    (0.59491544,11)
    (0.61252781,28)
    (0.63014019,19)
    (0.64775256,24)
    (0.66536493,31)
    (0.68297730,34)
    (0.70058967,42)
    (0.71820205,29)
    (0.73581442,35)
    (0.75342679,30)
    (0.77103916,36)
    (0.78865153,48)
    (0.80626391,53)
    (0.82387628,47)
    (0.84148865,48)
    (0.85910102,35)
    (0.87671340,47)
    (0.89432577,51)
    (0.91193814,55)
    (0.92955051,31)
    (0.94716288,35)
    (0.96477526,22)
    (0.98238763,36)
    (1.00000000,0)
  };
 
  \addplot[blue, solid, line width=2pt, forget plot] coordinates {
    (0.76178320,0) (0.76178320,55)
  };
  \addlegendimage{blue, solid, line width=2pt}
  \addlegendentry{Mean: 0.76}
 
  \addplot[red, dashed, line width=2pt, forget plot] coordinates {
    (0.79477455,0) (0.79477455,55)
  };
  \addlegendimage{red, dashed, line width=2pt, dash pattern=on 4pt off 2pt}
  \addlegendentry{Median: 0.79}
 
  \end{axis}
\end{tikzpicture}}
\caption{Distribution of OvR F1-score}
\label{fig:dist_ovr_f1}
\end{subfigure}
\hfill
\begin{subfigure}[b]{0.48\textwidth}
\centering
\resizebox{\linewidth}{!}{\begin{tikzpicture}
  \begin{axis}[
    title={},
    xlabel={Average Perceived Difficulty},
    ylabel={Average OvR-F1 Score},
    title style={font=\fontsize{11}{13}\selectfont},
    label style={font=\fontsize{9}{11}\selectfont},
    tick label style={font=\fontsize{9}{11}\selectfont},
    xmin=1, xmax=5,
    ymin=0.35, ymax=1.05,
    xtick={1,1.5,2,2.5,3,3.5,4,4.5,5},
    ytick={0,0.2,0.4,0.6,0.8,1.0},
    width=\linewidth,
    height=0.75\linewidth,
    grid=both,
    grid style={line width=0.4pt, draw=gray!30},
    xtick align=outside,
    ytick align=outside,
  ]
 
  \addplot[
    only marks,
    mark=*,
    mark size=1.5pt,
    mark options={fill=tealcolor, draw=black, line width=0.4pt},
  ] coordinates {
    (3.20000000,0.72829595)
    (4.00000000,0.84935438)
    (2.20000000,1.00000000)
    (2.33333333,0.89290935)
    (3.60000000,0.68835189)
    (3.80000000,0.63929795)
    (2.40000000,1.00000000)
    (3.00000000,0.70064513)
    (3.80000000,0.67938286)
    (2.66666667,0.92079331)
    (3.16666667,0.70544960)
    (3.00000000,0.77630836)
    (3.66666667,0.65176736)
    (1.66666667,0.88307210)
    (2.60000000,0.97777778)
    (3.33333333,0.79379457)
    (3.00000000,0.83293530)
    (3.33333333,0.83991325)
    (3.80000000,0.73017714)
    (3.16666667,0.63600165)
    (3.20000000,0.69844383)
    (3.80000000,0.74888382)
    (3.60000000,0.77809647)
    (2.80000000,0.86679514)
    (3.80000000,0.50315780)
    (3.00000000,0.75423777)
    (3.00000000,0.72754469)
    (2.60000000,0.82255972)
    (1.66666667,0.72299825)
    (3.40000000,0.87719107)
    (2.80000000,0.51328496)
    (3.20000000,0.62007933)
    (3.00000000,0.86141020)
    (3.66666667,0.88283608)
    (3.33333333,0.76028602)
    (3.33333333,0.65140854)
    (3.80000000,0.75073713)
    (3.66666667,0.71596293)
    (3.50000000,0.76470833)
    (3.66666667,0.92522821)
    (3.83333333,0.62272680)
    (3.60000000,0.68113141)
    (3.00000000,0.65609986)
    (3.00000000,0.76057856)
    (3.20000000,0.71954345)
    (3.66666667,0.67354179)
    (2.33333333,0.78877138)
    (3.66666667,0.74472983)
    (3.00000000,0.86325754)
    (2.80000000,0.82733552)
    (2.83333333,0.79945970)
    (2.50000000,0.78819859)
    (3.00000000,0.82108048)
    (2.66666667,0.89395315)
    (3.66666667,0.64825470)
    (2.42857143,0.85453181)
    (3.20000000,0.88998487)
    (3.00000000,0.79232525)
    (3.00000000,0.78244979)
    (3.33333333,0.75762532)
    (2.50000000,0.87795003)
    (2.83333333,0.82484056)
    (2.40000000,0.88546584)
    (2.83333333,0.86296723)
    (3.00000000,0.82688405)
    (3.40000000,0.85341471)
    (2.60000000,0.81993362)
    (2.66666667,0.84422738)
    (3.33333333,0.83280221)
    (2.50000000,0.72944768)
    (3.00000000,0.81909430)
    (3.50000000,0.78632987)
    (3.60000000,0.66111102)
    (2.83333333,0.58822071)
    (4.00000000,0.73756323)
    (3.80000000,0.74579468)
    (3.50000000,0.82110377)
    (3.80000000,0.66003496)
    (3.33333333,0.69782786)
    (3.50000000,0.63179819)
    (3.33333333,0.74814825)
    (3.66666667,0.68907719)
    (3.50000000,0.68898262)
    (3.20000000,0.94065934)
    (3.60000000,0.69350118)
    (3.50000000,0.57498262)
    (3.00000000,0.63240698)
    (3.16666667,0.66683879)
    (2.80000000,0.71545832)
    (3.00000000,0.76088084)
    (3.00000000,0.83884798)
    (3.66666667,0.72520147)
    (3.66666667,0.71934942)
    (3.33333333,0.62191030)
    (3.00000000,0.60982274)
    (3.16666667,0.84582331)
    (3.33333333,0.73935096)
    (3.80000000,0.77897542)
    (2.83333333,0.68799345)
    (3.20000000,0.76695786)
    (3.66666667,0.84154137)
    (3.83333333,0.43127651)
    (4.00000000,0.77089449)
    (4.00000000,0.91426640)
    (2.83333333,0.87268636)
    (2.83333333,0.60671116)
    (3.00000000,0.76870013)
    (3.60000000,0.57476781)
    (3.40000000,0.86337543)
    (2.66666667,0.86676431)
    (2.40000000,0.76611844)
    (3.00000000,0.91165655)
    (3.00000000,0.80778526)
    (3.16666667,0.87006895)
    (3.33333333,0.79217858)
    (2.80000000,0.73123398)
    (3.00000000,0.82660583)
    (3.00000000,0.78819073)
    (2.66666667,0.88356932)
    (2.40000000,0.93942587)
    (3.40000000,0.78054404)
    (3.33333333,0.77662818)
    (2.66666667,0.90058480)
    (3.00000000,0.91430499)
    (2.50000000,0.92361111)
    (3.16666667,0.81466186)
    (3.00000000,0.59331697)
    (3.33333333,0.73158163)
    (3.33333333,0.77625570)
    (2.60000000,0.84437573)
    (3.20000000,0.56189346)
    (2.66666667,0.81288826)
    (3.33333333,0.63022064)
    (3.20000000,0.71045153)
    (3.00000000,0.69551509)
    (3.50000000,0.63170323)
    (3.40000000,0.75751712)
    (2.40000000,0.95183946)
    (3.33333333,0.73466502)
    (3.60000000,0.72754868)
    (3.50000000,0.82711225)
    (3.00000000,0.78756912)
    (3.50000000,0.77382768)
    (3.00000000,0.80821224)
    (2.66666667,0.62402871)
    (4.00000000,0.66428472)
    (3.16666667,0.71495680)
    (3.40000000,0.72759612)
    (3.50000000,0.68334053)
    (3.75000000,0.64167982)
    (3.50000000,0.62293240)
    (3.80000000,0.71872176)
    (3.40000000,0.73207176)
    (2.66666667,0.71442490)
    (3.50000000,0.58602974)
    (3.20000000,0.87009734)
    (3.00000000,0.72313089)
    (3.50000000,0.74642890)
    (3.40000000,0.82395846)
    (3.50000000,0.70426542)
    (3.16666667,0.80644333)
    (3.50000000,0.61294598)
    (3.83333333,0.69928888)
    (3.00000000,0.89506140)
    (3.16666667,0.73254516)
    (3.00000000,0.62322040)
    (3.16666667,0.80364442)
    (2.83333333,0.91277495)
    (3.33333333,0.85172261)
    (3.40000000,0.72096784)
    (3.20000000,0.59157587)
    (3.00000000,0.76743772)
    (3.00000000,0.76274979)
    (2.60000000,0.93333333)
  };
 
  \end{axis}
\end{tikzpicture}}
\caption{OvR F1-score vs Difficulty}
\label{fig:ovr_f1_difficulty_corr}
\end{subfigure}
\caption{OvR F1-score related statistics}
\label{fig:combined_ovr_f1_plain}
\end{figure}

Regarding annotator agreement, for every change point \textit{P} flagged by an annotator, we check whether other annotators of the same timeseries agree on a change point within a 5-point radius. Out of 5781 annotated changes, 1275 (22.05\%) were not agreed upon by any other annotator, while 1415 (24.48\%) were agreed upon by all annotators. We define the One-versus-Rest F1 score (OvR F1) as the F1-score of an annotator's change points against the annotations of all other annotators of the same timeseries, serving as a measure of collective agreement. Figure \ref{fig:dist_ovr_f1} shows that most tasks score closer to 1 than to 0, confirming general agreement among annotators, with a mean OvR F1 of 0.76 and a median of 0.79. 
Figure \ref{fig:ovr_f1_difficulty_corr} further shows that harder timeseries tend to yield lower OvR F1 scores (Pearson \textit{r} = -0.42), suggesting that perceived difficulty negatively correlates with annotation agreement.
Overall, a mean OvR F1 of 0.76 and the strong correlation between difficulty and agreement collectively indicate that the annotation data is of sufficient quality for our evaluation purposes: annotators reliably agree on unambiguous change points, and disagreements are largely confined to inherently difficult time series rather than reflecting systematic inconsistency.

\begin{table}
  \caption{Summary statistics of the annotation experiment}
  \label{tab:general_stat}
  \centering
  \begin{tabular}{@{} l l r @{}}
    \toprule
    & \textbf{Metric} & \textbf{Value} \\
    \midrule
    \multirow{3}{*}{\textit{Scope}}
      & Annotators                  & 11     \\
      & Time series                 & 174    \\
      & Total tasks                 & 975    \\
    \cmidrule(l){2-3}
    \multirow{3}{*}{\textit{Per-task statistics}}
      & Avg.\ change points flagged & 6.3    \\
      & Avg.\ duration              & 323\,s \\
      & Median duration             & 65\,s  \\
    \cmidrule(l){2-3}
    \multirow{2}{*}{\textit{Issues}}
      & Platform issues             & 0      \\
      & External issues             & 4      \\
    \bottomrule
  \end{tabular}
\end{table}

\subsection{Offline Methods' Results}
\label{subsec:offline}

Table~\ref{tab:cpd_results_offline_no_variance} reports F1-score, precision, and recall for all offline methods across the three groups of configurations. 
\currentmozillamethod with an F1-score of 70.6\%, a precision of 99.2\%, and a recall of 57.7\% serves as the reference point, with cells highlighted in green indicating values that exceed its corresponding metric. 
We also include the \texttt{Zero} baseline, which flags no change points, to help characterize the dataset. While this method trivially achieves perfect precision, its recall is notably non-zero as in 25.9\% of the time series, at least one annotator indicates that no changes are present.

\begin{table}[t!]
\centering
\caption{Performance Metrics for Offline Methods. We highlight in \colorsq{green!40} the results that surpass \currentmozillamethod on the corresponding evaluation metric.
  }
\label{tab:cpd_results_offline_no_variance}
\resizebox{\textwidth}{!}{%
\begin{tabular}{l ccc|ccc|ccc}
\toprule
& \multicolumn{3}{c|}{Default Configuration} & \multicolumn{3}{c|}{Best Configuration} & \multicolumn{3}{c}{Upper-bound Performance} \\
\cmidrule(lr){2-4} \cmidrule(lr){5-7} \cmidrule(lr){8-10}
Method & F1 & Precision & Recall & F1 & Precision & Recall & F1 & Precision & Recall \\

\midrule
Zero (Baseline) & 0.407      & 1.000    & 0.268 & -- & -- & -- & -- & -- & -- \\
\currentmozillamethod & 0.706 & 0.992 & 0.577 & -- & -- & -- & -- & -- & -- \\

\midrule

BinSeg        & 0.675 & 0.859 & \good{0.594}  & 0.699 & 0.749 & \good{0.711}  & \good{0.809} & \good{1.000} & \good{1.000} \\

CPNP          & 0.694 & 0.634 & \good{0.819} & \good{0.724} & 0.668 & \good{0.840} & \good{0.754} & 0.701 & \good{1.000} \\

KCPA          & 0.618 & 0.864 & 0.513 & \good{0.728} & 0.697 & \good{0.830} & \good{0.835} & 0.957 & \good{0.844} \\

MongoDB       & \good{0.734} & 0.789 & \good{0.731} & \good{0.740} & 0.763 & \good{0.760} & \good{0.774} & 0.828 & \good{0.779} \\

PELT          & 0.644 & 0.750 & \good{0.626} & 0.660 & 0.703 & \good{0.692} & \good{0.752} & 0.810 & \good{1.000} \\

RFPOP         & 0.701 & 0.613 & \good{0.884} & 0.703 & 0.616 & \good{0.884} & \good{0.714} & 0.629 & \good{0.949} \\

SegNeigh      & 0.659 & 0.792 & \good{0.604} & 0.659 & 0.703 & \good{0.691} & \good{0.787} & \good{1.000} & \good{1.000} \\

WBS           & 0.513 & 0.400 & \good{0.864} & 0.666 & 0.615 & \good{0.789} & \good{0.730} & 0.698 & \good{1.000} \\

AMOC          & 0.541 & 0.922 & 0.406 & 0.542 & 0.922 & 0.407 & 0.582 & \good{1.000} & 0.433 \\

\bottomrule
\end{tabular}
}
\end{table}

\textit{\textbf{Across the board, CPD methods are more sensitive to performance changes than \currentmozillamethod.}} Under the \textit{Default} configuration, nearly all methods outperform \currentmozillamethod's recall of 57.7\%, meaning they miss fewer true regressions. 
However, this comes at a steep precision cost as most methods issue substantially more false alerts, with precision values as low as 40\% for \texttt{WBS} and 61\% for \texttt{RFPOP}. This could reflect a fundamental behavioral difference which is that CPD methods are designed to be sensitive to data shifts, whereas \currentmozillamethod is deliberately calibrated for precision. 
In the aggregated F1-score, only \texttt{MongoDB} improves the F1-score in this setting achieving 73.4\%, capturing 73.1\% of all annotated regressions while keeping false alerts rate at 21\%.

\textit{\textbf{There are real performance gains when methods are tuned.}}  
The gains of moving from \textit{Default} to \textit{Best} configurations are substantial and consistent across all methods, ranging from modest improvements to substantial increases in performance. For example, \texttt{WBS}'s F1-score improves from 51.3\% to 66.6\%, and \texttt{KCPA}'s F1-score improves from 61.8\% to 72.8\%. 
Once tuned, three methods achieve an overall improvement over \currentmozillamethod as \texttt{CPNP}, \texttt{KCPA}, and \texttt{MongoDB} methods achieve an F1-score of 72.4\%, 72.8\%, and 74\% respectively. 
Among these, \texttt{MongoDB} is the most robust of the evaluated methods and outperforms \currentmozillamethod's F1-score in both \textit{Default} and \textit{Best} configurations.  
Tuning generally improves recall at the cost of precision, as five out of nine methods see a drop in precision, notably with \texttt{KCPA}, which shows the most pronounced drop, moving from 86.4\% to 69.7\% from \textit{Default} to \textit{Best} configuration. Concerning recall, all methods except for \texttt{WBS} see an increase. So, performance generally improves by having a better precision at the slight cost of a lower recall.

\textit{\textbf{Upper-bound performance of CPD methods shows significant gains in recall, but most fail to reach the precision of \currentmozillamethod}}. 
Under \textit{Upper-Bound Performance}, which selects the best possible configuration per time series, almost all methods surpass \currentmozillamethod, with \texttt{KCPA} reaching an F1-score of 83.5\% and \texttt{BinSeg} reaching 80.9\%. All metrics increase going from \textit{Best} configuration to \textit{Upper-Bound Performance}, with an improvement of 10.15\% for F1-score, 18.46\% for precision, and 21.12\% for recall.
It is worth noting that several methods achieve a precision or recall of 100\% under \textit{Upper-Bound Performance}: a perfect recall as seen in methods such as \texttt{BinSeg}, and \texttt{CPNP} reflects over-reporting, where the method flags so many change points that all true regressions are inevitably captured. Conversely, a perfect precision which is as seen in \texttt{AMOC}, \texttt{BinSeg}, and \texttt{SegNeigh} reflects under-reporting, where so few points are flagged that every detection happens to be correct.

\summarybox{How effective are Offline CPD methods in detecting performance changes?}{Offline CPD methods show greater sensitivity to performance changes, leading to better recall (fewer missed alerts) at the cost of worse precision (more false alerts).
When tuned to performance measurements, \texttt{CPNP}, \texttt{KCPA}, and \texttt{MongoDB}, show better overall performance than \currentmozillamethod, improving the F1-score by 2.55\%, 3.12\%, and 4.82\% respectively.}

\subsection{Online Methods' Results}
\label{subsec:online}

We present the results of the online methods evaluation in Table~\ref{tab:cpd_results_online_no_variance}.
Once again, the main baseline is the performance of \currentmozillamethod, and we also include the \texttt{ODummy} baseline, which flags change points at fixed intervals of 300 time steps.
The method, hence, serves as a sanity check for online methods operating without any statistical grounding, yielding an F1-score of 26.3\%.

\begin{table}[t!]
\centering
\caption{Performance Metrics for Online Methods. For colour interpretation, see Table~\ref{tab:cpd_results_offline_no_variance}. We highlight in \colorsq{green!40} the results that surpass \currentmozillamethod on the corresponding evaluation metric.}
\label{tab:cpd_results_online_no_variance}
\resizebox{\textwidth}{!}{%
\begin{tabular}{l ccc|ccc|ccc}
\toprule
& \multicolumn{3}{c|}{Default Configuration} & \multicolumn{3}{c|}{Best Configuration} & \multicolumn{3}{c}{Upper-Bound Performance} \\
\cmidrule(lr){2-4} \cmidrule(lr){5-7} \cmidrule(lr){8-10}
Method & F1 & Precision & Recall & F1 & Precision & Recall & F1 & Precision & Recall \\
\midrule

ODummy (Baseline) & 0.263 & 0.325 & 0.287 & -- & -- & -- & -- & -- & -- \\

\currentmozillamethod & 0.706 & 0.992 & 0.577 & -- & -- & -- & -- & -- & -- \\
\midrule

ADWIN         & 0.272 & 0.285 & 0.310 & 0.299 & 0.384 & 0.285 & 0.485 & 0.617 & 0.472 \\

CVMWIN        & 0.203 & 0.148 & 0.497 & 0.225 & 0.155 & \good{0.632} & 0.319 & 0.245 & \good{0.770} \\

BOCPD         & 0.633 & 0.577 & \good{0.803} & 0.671 & 0.731 & \good{0.677} & \good{0.799} & 0.939 & \good{0.987} \\

CUSUM         & 0.220 & 0.165 & 0.520 & 0.313 & 0.370 & 0.331 & 0.518 & 0.622 & \good{0.664} \\

EWMA          & 0.407 & \good{1.000} & 0.268 & 0.408 & \good{1.000} & 0.269 & 0.510 & \good{1.000} & 0.461 \\

KSWIN         & 0.281 & 0.286 & 0.334 & 0.302 & 0.357 & 0.304 & 0.455 & 0.574 & 0.475 \\

Page Hinkley  & 0.291 & 0.347 & 0.288 & 0.387 & 0.682 & 0.295 & 0.525 & 0.845 & 0.490 \\

Shewhart      & 0.296 & 0.332 & 0.459 & 0.323 & 0.405 & 0.405 & 0.406 & 0.511 & \good{0.667} \\

SPRT          & 0.205 & 0.147 & 0.512 & 0.226 & 0.137 & \good{0.979} & 0.249 & 0.166 & \good{0.979} \\

\bottomrule
\end{tabular}
}
\end{table}

\textit{\textbf{Out of the box, most online methods surpass the trivial baseline.}} Under the \textit{Default} configuration, most methods perform better than the \texttt{ODummy} baseline in terms of F1-score with the only three performing worse with F1-scores as low as 22\% for CUSUM, 20.5\% for SPRT, and 20.3\% for CVMWIN. However, F1-score values of most of the other methods remain close to that of the dummy baseline. This is not a sign of fundamental weakness in the underlying statistical tests, but rather a consequence of high sensitivity to parameter initialization, a pattern consistent with what we observe in offline methods. \texttt{BOCPD} is the clear exception, reaching an F1-score of 63.3\% without any tuning and a recall of 80.3\% already surpassing \currentmozillamethod's recall, meaning it catches more true regressions out of the box at the cost of lower precision.

\textit{\textbf{Tuning improves results considerably, but no online method closes the gap with \currentmozillamethod.}} After tuning, all methods improve over their \textit{Default} results, with some methods seeing dramatic gains such as \texttt{Page Hinkley} and \texttt{CUSUM}. 
Yet despite these gains, no online method surpasses \currentmozillamethod's F1-score of 70.6\% under \textit{Best} configuration. \texttt{BOCPD} remains the strongest candidate with an F1-score of 67.1\%, missing fewer annotated regressions as recall is 67.7\%, but still issuing 26.9\% of false alerts. However, \texttt{BOCPD} remains substantially lower than that of \currentmozillamethod.
Contrary to the offline methods, tuning online methods yields better precision at the cost of recall which significantly decreases for four out of nine evaluated methods, namely for \texttt{BOCPD}, \texttt{ADWIN}, \texttt{Page Hinkley}, and \texttt{Shewhart}.

\textit{\textbf{The Upper-Bound performance reveals a ceiling on tuning as even with optimal parameters, online methods cannot match \currentmozillamethod.}} Under \textit{Upper-Bound Performance}, only one online method surpasses \currentmozillamethod's F1-score, which is \texttt{BOCPD} as it reaches its highest achievable F1-score of 79.9\%, yet this remains below the best offline \textit{Upper-Bound Performance} result ,which is \texttt{KCPA} at an 83.5\% F1-score, indicating that the performance ceiling for online methods lies below that of their offline counterparts. This means that additional tuning effort yields diminishing returns. It is also worth noting that \texttt{ADWIN}, \texttt{KSWIN}, and \texttt{CVMWIN} retain some retrospective capacity by attributing detections to earlier positions within an observation window.

\textit{\textbf{EWMA method presents an outlier.}} Upon investigating \texttt{EWMA}'s experiments results, it either produces no alerts at all or simultaneously floods with false positives and misses many true regressions, yielding F1-scores lower than issuing no alerts whatsoever. The configurations that avoid this erratic behavior are inherently conservative, explaining the perfect precision but low recall consistently observed across all configurations. This also makes \texttt{EWMA} an outlier in terms of variance across configurations. One plausible explanation is that the hyperparameter space 
itself, despite covering 108 distinct configurations as per Table~\ref{tab:cpd_methods_hyperparam}, does not expose a regime in which \texttt{EWMA} behaves neither too conservatively nor too erratically, pointing to a fundamental sensitivity of the method to hyperparameter choices beyond the ranges explored.

\summarybox{How effective are online CPD methods for identifying performance regressions?}{Online CPD methods prioritize issuing alerts in real time, and this shows an inherent limitation in their effectiveness.
Most evaluated online CPD methods have lower precision and recall, compared to~\currentmozillamethod.
The exception is \texttt{BOCPD}, which shows comparable out-of-the-box and tuned performance, showing an improvement of 17.33\% in recall at a cost of 26.31\% in precision loss when fine-tuned.}

\subsection{Hybrid Methods' Results}
\label{subsec:hybrid}

We report in Table~\ref{tab:cpd_results_stats_no_variance} the results of all hybrid methods.

\begin{table}[t!]
\centering
\caption{Performance Metrics for Hybrid Methods. For colour interpretation, see Table~\ref{tab:cpd_results_offline_no_variance}. We highlight in \colorsq{green!40} the results that surpass \currentmozillamethod on the corresponding evaluation metric.}
\label{tab:cpd_results_stats_no_variance}
\resizebox{\textwidth}{!}{%
\begin{tabular}{l ccc ccc ccc}
\toprule
Method & \multicolumn{3}{c}{Default Configuration} & \multicolumn{3}{c}{Best Configuration} & \multicolumn{3}{c}{Upper-Bound Performance} \\
\cmidrule(lr){2-4} \cmidrule(lr){5-7} \cmidrule(lr){8-10}
 & F1-score & Precision & Recall & F1-score & Precision & Recall & F1-score & Precision & Recall \\

\midrule
\currentmozillamethod & 0.706 & 0.992 & 0.577 & -- & -- & -- & -- & -- & -- \\
\midrule

CVM & 0.411 & 0.295 & \good{0.908} & 0.622 & 0.569 & \good{0.792} & \good{0.749} & 0.729 & \good{0.987} \\
MWU & 0.403 & 0.287 & \good{0.902} & 0.643 & 0.708 & \good{0.642} & \good{0.777} & 0.831 & \good{0.983} \\
KS & 0.396 & 0.277 & \good{0.901} & 0.631 & 0.627 & \good{0.711} & \good{0.757} & 0.739 & \good{0.981} \\
LEVENE & 0.369 & 0.335 & 0.544 & 0.493 & 0.567 & 0.517 & 0.658 & 0.762 & \good{0.854} \\
WELCH & 0.451 & 0.343 & \good{0.885} & 0.671 & 0.647 & \good{0.787} & \good{0.750} & 0.723 & \good{0.978} \\

\bottomrule
\end{tabular}
}
\end{table}

\textit{\textbf{Out of the box, hybrid methods have better recall but are less precise than \currentmozillamethod.}} Under \textit{Default} configuration, nearly all methods surpass \currentmozillamethod's recall of 57.7\%, with \texttt{CVM}, \texttt{MWU}, and \texttt{KS} exceeding 90\%. However, precision drops sharply that are as low as 27.7\% (\texttt{KS}), and no method matches \currentmozillamethod's F1-score of 70.6\%. \texttt{LEVENE} is the weakest overall, failing to beat \currentmozillamethod on either recall or precision.

\textit{\textbf{Tuning consistently improves F1-score, but methods still underperform, if compared to \currentmozillamethod.}} All methods improve after tuning, with some precision gains with \texttt{MWU} rising from 28.7\% to 70.8\%. However, none surpass \currentmozillamethod's F1-score. Methods remain too aggressive in flagging change points for competitive overall performance.

\textit{\textbf{Upper-Bound Performance reveals meaningful calibration headroom.}} Several methods exceed \currentmozillamethod under the upper-bound performance, with \texttt{MWU} reaching 77.7\% and \texttt{CVM} reaching a 74.9\% F1-score. The gap between \textit{Best} and \textit{Upper-Bound} configurations suggests that better per-series calibration could substantially close the precision gap.

\summarybox{How effective are hybrid CPD methods for identifying performance regressions?}{Hybrid methods detect more true regressions than \currentmozillamethod, but their precision is too low for practical use without tuning. The fine-tuning results suggest this is a calibration challenge, not a fundamental limitation.}

\subsection{Ensemble Voting Results}
\label{subsec:hybrid_methods_voting_results}

\noindent\textbf{Approach.} The results of the individual CPD methods reveal a clear tradeoff: most methods tend to produce a high false positive rate with an acceptable false negative rate, whereas \currentmozillamethod exhibits the opposite issue.
This complementary behaviour motivates the design of an ensemble voting framework that aggregates the outputs of multiple methods into a collective decision, enabling more robust detection than any single method alone. 
We apply this framework to all three families of methods evaluated in this section: offline, online, and hybrid.

To construct the voting ensemble, we select the four best-performing offline methods based on their F1-score under the \textit{Best} configuration, excluding lower-performing methods to avoid diluting the ensemble's detection quality. 
We adopt the same rationale for the online methods. Eventually, the offline methods participating in the voting are \texttt{BinSeg}, \texttt{CPNP}, \texttt{KCPA}, and \texttt{MongoDB}, and the online methods participating are \texttt{BOCPD}, \texttt{EWMA}, \texttt{Page Hinkley}, and \texttt{Shewhart}.
For both offline and online methods, we apply a voting ensemble with equal weights, a.k.a. \textit{Ensemble} voting. 
In the ensemble of hybrid methods, where the complementarity with \currentmozillamethod is most pronounced, we also experiment with a mode of \textit{Mozilla or Ensemble} strategy which is defined as follows: \currentmozillamethod performs well in terms of precision but not in terms of recall, which means that whenever it flags an alert, it is most likely a true alert, but it misses on several alerts that should have been detected. That gap of these undetected alerts is filled by a voting ensemble encompassing all of the other hybrid methods with equal weights.

\noindent\textbf{Results.}
Tables ~\ref{tab:voting_strategies_results_offline_online} and ~\ref{tab:voting_strategies_results_hybrid} report the results of the voting strategies across all method families under increasing consensus thresholds, with \currentmozillamethod serving as the baseline reference. Table~\ref{tab:voting_strategies_results_hybrid} presents the \textit{Ensemble} and \textit{Mozilla or Ensemble} strategies applied to hybrid methods, where the latter gives \currentmozillamethod an unconditional vote before applying the consensus threshold to the remaining methods. Table~\ref{tab:voting_strategies_results_offline_online} presents the \textit{Ensemble} voting strategy applied independently to the offline and online method families.

\begin{table}
\centering
\caption{Voting strategies results using offline \& online methods. We highlight in \colorsq{green!40} the results that surpass \currentmozillamethod on the corresponding evaluation metric. The best F1-score within each voting strategy is in \textbf{bold}.
}
\label{tab:voting_strategies_results_offline_online}
\begin{tabular}{ll ccc c ccc}
\toprule
& & \multicolumn{3}{c}{Offline Methods} & & \multicolumn{3}{c}{Online Methods} \\
\cmidrule(lr){3-5} \cmidrule(lr){6-9}
Strategy & Consensus ($C$) & F1 & Precision & Recall & & F1 & Precision & Recall \\
\midrule

Baseline & \currentmozillamethod & 0.706 & 0.992 & 0.577 & & 0.706 & 0.992 & 0.577 \\
\midrule
\multirow{4}{*}{Ensemble}& C=1  & 0.681 & 0.582 & \good{0.888} & & 0.477 & 0.411 & \good{0.702} \\
& C=2 & \good{0.782} & 0.770 & \good{0.830} & & \textbf{0.491} & 0.812 & 0.386 \\
& C=3 & \textbf{0.784} & 0.864 & \good{0.753} & & 0.413 & 0.977 & 0.275 \\
& C=4 & \good{0.754} & 0.940 & \good{0.661} & & 0.408 & \good{1.000} & 0.269 \\
\bottomrule
\end{tabular}
\end{table}

\begin{table}[t!]
\centering
\caption{Voting strategies results using hybrid methods. We highlight in \colorsq{green!40} the results that surpass \currentmozillamethod on the corresponding evaluation metric. Voting strategies results. The best F1-score within each voting strategy is in \textbf{bold}.
}

\label{tab:voting_strategies_results_hybrid}
\begin{tabular}{l l ccc}
\toprule
& & \multicolumn{3}{c}{Results} \\
\cmidrule(lr){3-5}
Strategy & Consensus ($C$) & F1 & Precision & Recall \\
\midrule
Baseline & \currentmozillamethod & 0.706 & 0.992 & 0.577 \\
\midrule

\multirow{6}{*}{Ensemble}
& C=1  & 0.592 & 0.493 & 0.863 \\
& C=2 & \good{0.724} & 0.718 & \good{0.795} \\
& C=3 & \good{\textbf{0.768}} & 0.868 & \good{0.726} \\
& C=4 & \good{0.747} & 0.950 & \good{0.645} \\
& C=5 & 0.680 & 0.983 & 0.552 \\
& C=6 & 0.577 & \good{0.999} & 0.434 \\

\midrule

\multirow{5}{*}{\shortstack{Mozilla or \\ Ensemble}}
& C=1  & 0.593 & 0.494 & \good{0.865} \\
& C=2 & \good{0.722} & 0.706 & \good{0.802} \\
& C=3 & \good{\textbf{0.772}} & 0.855 & \good{0.738} \\
& C=4 & \good{0.763} & 0.944 & \good{0.669} \\
& C=5 & \good{0.726} & 0.979 & \good{0.606} \\

\bottomrule
\end{tabular}
\end{table}

\textit{\textbf{Offline methods form the strongest ensemble, surpassing \currentmozillamethod without requiring its participation.}} In Table~\ref{tab:voting_strategies_results_offline_online}, each row corresponds to a consensus level $C$, where a point is flagged only if at least $C$ methods agree. Lower $C$ values favor recall while higher ones favor precision. The offline ensemble peaks at an F1-score of 78.4\% at C=3, surpassing \currentmozillamethod's F1-score of 70.6\% by 7.8 percentage points, and achieved without including \currentmozillamethod in the ensemble at all. Precision and recall are well-balanced at this threshold, scoring 86.4\% and 75.3\% respectively, reflecting a meaningful improvement over any individual offline method. As the threshold increases to C=4, precision continues to rise to 94.0\% but recall drops to 66.1\%, confirming the expected tradeoff at high consensus thresholds. The ensemble benefits from the diversity of the four selected methods, where their disagreements filter out incidental detections and their agreements reinforce genuine regressions. It is important to note that offline ensembles and hybrid ones have comparable results, both peaking in terms of F1-score at 78.4\% and 77.2\% respectively.

\textit{\textbf{Online methods do not form an effective ensemble.}} The online voting ensemble peaks at a modest F1-score of 49.1\% at C=2, far below \currentmozillamethod. Even at C=1, the ensemble reaches only 47.7\%, lower than any offline configuration. At C=4, precision reaches 100\% but recall collapses to 26.9\%, reflecting an under-reporting behavior.

\textit{\textbf{Hybrid methods combined with \currentmozillamethod through voting produce the best balance between precision and recall.}} Table~\ref{tab:voting_strategies_results_hybrid} shows that both voting strategies surpass \currentmozillamethod at moderate thresholds, with \textit{Ensemble voting (C=3)} peaking at F1-score of 76.8\% and \textit{Mozilla or Ensemble (C=3)} at an F1-score of 77.2\%. As the agreement threshold increases, precision rises and recall falls monotonically, with F1-score peaking before declining as the precision gain no longer compensates for the recall loss. \textit{Mozilla or Ensemble} consistently outperforms \textit{Ensemble} voting at the same threshold, and the advantage becomes most apparent at high thresholds as \textit{Mozilla or Ensemble (C=5)} still substantially outperforms \textit{Ensemble (C=6)} by 14.9 percentage points in F1-score. This confirms that preserving \currentmozillamethod's high-precision detections unconditionally prevents the steep recall collapse that \textit{Ensemble} voting suffers at high consensus thresholds.

\textit{\textbf{Ensemble methods help overcome the limitations of individual CPD methods.}}
Taken in isolation, hybrid methods do not surpass the best offline method \texttt{MongoDB} with an F1-score of 74.0\% after tuning, and their precision remains a persistent weakness. The ensemble resolves this by combining the high recall of hybrid methods with the high precision of \currentmozillamethod through voting, achieving a balance that no individual method reaches. Similarly, the offline ensemble demonstrates that aggregating diverse methods filters noise effectively even without \currentmozillamethod's participation.

\summarybox{How effective is combining CPD methods in identifying performance regressions?}{Ensemble voting surpasses individual methods across offline and hybrid families, with offline ensemble peaking at 78.4\% F1-score and hybrid ensemble paired with \currentmozillamethod at 77.2\%. Online ensemble remains ineffective regardless of the consensus threshold.}

\section{\textbf{Practitioner Survey Validation}}
\label{sec:industrial_validation}

We present the method and results of our practitioner validation study, which consists of a workshop and a survey study to present the results of our experimental comparison of CPD methods to Mozilla's performance engineering team. 
Participants are asked to participate in a survey study that has two main complementary goals:

\begin{enumerate}
    \item \textbf{Validate parameters of the experimental study:} In our evaluation study, we assume two important parameters: (a)~the applicability of the tolerance margin of $M=5$ revisions used when matching predicted change points to ground-truth annotations, and (b)~the importance of false positives and false negatives in the CPD result comparison through the F1 score. 
    \item \textbf{Validating the proposed solutions:}  We present the performance of \currentmozillamethod alongside the hybrid CPD methods and voting strategies.
    Given the comparable performance of the offline and hybrid voting methods, we opt to proceed with hybrid methods as they are easier to integrate and follow the same hybrid structure of \currentmozillamethod.
    Therefore, our survey focuses on validating the feasibility of the hybrid solutions, focusing on the two most promising voting strategies. 
\end{enumerate}

We present the core findings of our study during a scheduled Mozilla team meeting to a group of 21 practitioners. 
The presentation cover four topics: (i)~\currentmozillamethod and its limitations, (ii)~the practitioner-annotation baseline and evaluation framework, (iii)~the hybrid CPD methods and their results, and (iv)~the two voting strategies. To ensure broad and informed participation, we prepare two supporting artifacts: the structured follow-up survey, circulated through a 
channel with \textbf{91 members}, which includes screenshots and descriptions from the presentation, and a report detailing the key concepts and results.

\subsection{Survey Instrument Design}
\label{subsec:survey_instrument_design}

\begin{table}
\centering
\footnotesize
\setlength{\tabcolsep}{5pt}
\caption{Summary of the follow-up survey questions and answer formats.}
\begin{tabularx}{\columnwidth}{p{1.2cm} X p{2.1cm}}
\toprule
\textbf{Category} & \textbf{Question} & \textbf{Answer Type} \\
\midrule

\multirow{6}{*}{Background}
& \textbf{Q1.} Do you consent to participate in this follow-up survey?
& Tick box \\

& \textbf{Q2.} What roles do you have at Mozilla?
& Free text \\

& \textbf{Q3.} How many years have you been working in software development?
& Free text \\

& \textbf{Q4.} How many years have you been working in a performance-related area at Mozilla?
& Free text \\

& \textbf{Q5.} For how many years have you been using \textit{Perfherder}?
& Free text \\

& \textbf{Q6.} Did you take part in the workshop presentation (Feb 19, 2026)?
& Yes / No \\

\midrule
\multirow{6}{*}{\shortstack{Research \\ assumptions \\ validation}}
& \textbf{Q7.} The tolerance margin of 5 points is acceptable when evaluating detection methods.
& Likert scale (1--5) \\

& \textbf{Q8.} (Optional) Please elaborate on the reasons for your answer above.
& Free text \\

& \textbf{Q9.} A reliable detection system should minimize false and missing alerts equally.
& Likert scale (1--5) \\

& \textbf{Q10.} Minimizing false alerts should be prioritized due to validation effort.
& Likert scale (1--5) \\

& \textbf{Q11.} Minimizing missing alerts should be prioritized due to user impact.
& Likert scale (1--5) \\

& \textbf{Q12.} (Optional) Please elaborate on the reasons for your answer above.
& Free text \\

\midrule

\multirow{7}{*}{\shortstack{Result \\ validation}}
& \textbf{Q13.} The current method's performance in minimizing false alerts is acceptable.
& Likert scale (1--5) \\

& \textbf{Q14.} The current method's performance in minimizing missing alerts is acceptable.
& Likert scale (1--5) \\

& \textbf{Q15.} (Optional) Please elaborate on your assessment of the current method's performance.
& Free text \\

& \textbf{Q16.} How do you grade the overall performance of the statistical methods presented?
& MOS scale \\

& \textbf{Q17.} How do you grade the overall performance of the voting strategies presented?
& MOS scale \\

& \textbf{Q18.} What group of techniques shows adequate performance for deployment on \textit{Perfherder}?
& Multiple selection \\

& \textbf{Q19.} (Optional) Please elaborate on your answer above.
& Free text \\

\midrule

\multirow{2}{*}{Other}
& \textbf{Q20.} Would you like to suggest, question or comment anything related to this study?
& Free text \\

& \textbf{Q21.} Would you like to be notified about a pre-print of the study?
& Free text (email) \\

\bottomrule
\end{tabularx}
\label{tab:survey_summary}
\end{table}

The survey comprises 21 questions organized into four categories (Table~\ref{tab:survey_summary}), with an estimated completion time of 10 to 15 minutes.
The Background category (Q1--Q6) collects demographic information and consent.
The Research Assumptions Validation category (Q7--Q12) asks practitioners to rate two key assumptions: the tolerance margin and the equal weighting of false and missing alerts.
The Result Validation category (Q13--Q19) asks practitioners to grade \currentmozillamethod, the CPD methods, and voting strategies.
The Other category (Q20--Q21) collects open-ended feedback and optional contact information.
Each category includes an optional free-text field for elaboration, and only the consent question is mandatory.

The survey employs three rating methodologies:
\begin{itemize}
    \item \textbf{Likert-scale agreement questions} for statements about research assumptions and the acceptability of \currentmozillamethod performance~\cite{Likert1932}.
    \item \textbf{Mean Opinion Score (MOS)-style quality-assessment questions}, where practitioners quantify their perception of each technique on a five-point scale (1 corresponding to \textit{Poor} to 5 corresponding to \textit{Excellent}), with the mean score across participants serving as the final quality indicator~\cite{kirkland2023stuck}.
    In this survey, we asked participants to rate the CPD methods and voting strategies.
    \item \textbf{Optional free-text boxes} accompany each question group so that respondents can elaborate on their ratings.
\end{itemize}

\subsection{Survey Demographics}
\label{subsec:survey_demographics}

Participants have one week to fill out the survey, and we receive responses from 12 practitioners, nine of whom attended the presentation.
Table~\ref{tab:survey_participants} presents their demographic profiles.
On average, respondents have 11.9~years of software engineering experience, 4.3~years of Mozilla performance engineering experience, and 4.9~years of \textit{Perfherder} experience.
The participant pool mirrors that of the annotation study (Table~\ref{tab:demographics}), composed predominantly of Performance Sheriffs and Engineers, the practitioners who interact daily with \textit{Perfherder}.

\begin{table}
\centering
\renewcommand{\arraystretch}{0.9}
\caption{Demographic information about the survey participants.}
\begin{tabular}{llrrr}
\toprule
& & \multicolumn{3}{c}{\textbf{Years of experience}}\\ 
\textbf{ID} & \textbf{Role} & \textbf{Engineering} & \textbf{Performance} & \textbf{PerfHerder} \\
\midrule
P1 & Performance Tech Lead & 20 & 8 & 16 \\

P2 & Performance Engineer & 27 & 7.5 & 7.5 \\

P3 & Engineering Manager & 20 & 7 & 7 \\

P4 & Performance Tools Engineer & 7 & 1 & 0 \\

P5 & JavaScript Engine Developer & 12 & 7 & 7 \\

P6 & Engineer & 12 & 0 & 0 \\

P7 & Performance Sheriff \& Performance Test Engineer & 15 & 4 & 4 \\

P8 & Performance Sheriff & 7 & 5 & 5 \\

P9 & Performance Sheriff & 6 & 5 & 5 \\

P10 & Performance Sheriff and Performance Test & 7 & 2 & 2 \\

P11 & performance test/tools engineer & 8 & 4 & 4 \\

P12 & Performance Test Engineer & 2 & 1 & 1 \\

\midrule
    & \textbf{Average} & \textbf{11.9} & \textbf{4.3} & \textbf{4.9} \\
\bottomrule
\end{tabular}
\label{tab:survey_participants}
\end{table}

\begin{figure}
    \centering
    \includegraphics[width=1.0\columnwidth]{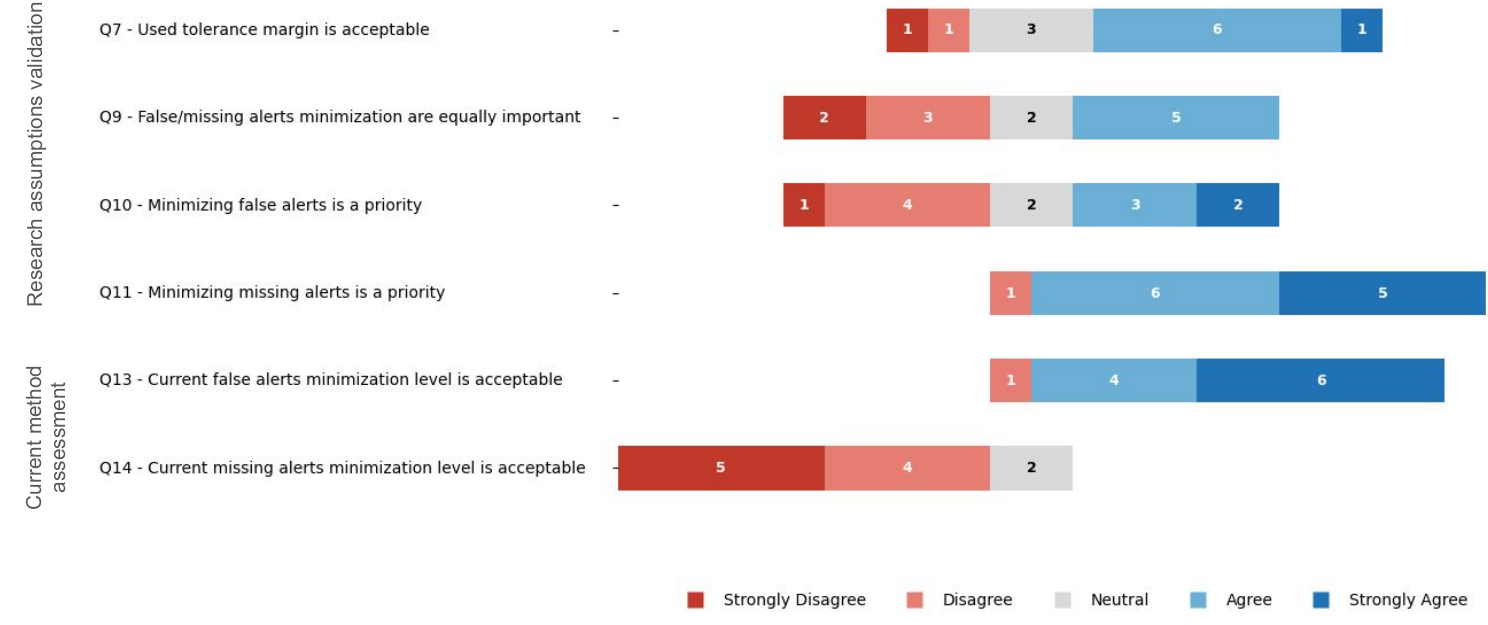}
    \caption{Likert scale questions results}
    \label{fig:likert_scale_agreement}
\end{figure}

\subsection{Validating Research Assumptions}
\label{subsec:validating_research_assumptions}

\noindent
\textbf{Validity of the Tolerance Margin.}
We use the tolerance margin of $M=5$ to evaluate the predicted change points against ground-truth annotations. 
\textit{Q7} in Figure~\ref{fig:likert_scale_agreement} shows that the majority of respondents find this margin acceptable: 7 out of 12 agree or strongly agree that a margin of five points is acceptable in practice.
Most practitioners consider a window of 5 revisions to be a reasonable operational tolerance for evaluating detection accuracy within \textit{Perfherder}'s workflow, and that a smaller margin would be considered overly strict and potentially unfair in practice.
\practitionerquote{A sensitivity analysis with different margin values would help justify this choice. But in the given example, even human annotators could disagree. Expecting a perfect match would be too harsh.}{P11}

\noindent
\textbf{Validating Penalization-Related Assumptions}
The second assumption concerns the equal weighting of false positives (false alerts) and false negatives (missing alerts) in the F1 score.
Practitioners express more nuanced views on this assumption, captured through \textit{Q9} to \textit{Q11}.

Respondents are split when asked to agree that both false alerts and missing alerts are equally important (Q9).  
This balance suggests that practitioners hold no strong consensus on this trade-off, with some viewing it as context-dependent. 
However, when asked to assess each error type independently, responses to \textit{Q10} and \textit{Q11} reveal a more pronounced asymmetry in practitioners' priorities. 
Minimizing missing alerts (\textit{Q11}) is strongly endorsed, with eleven out of twelve respondents agreeing or strongly agreeing with the emphasis on higher technique recall.
\practitionerquote{If a compromise needs to be done between the two, there should be no missed alerts, since that is what finally impacts the end user.}{P7}
Another practitioner introduces an important nuance, distinguishing between the severity of individual regressions:
\practitionerquote{Missing a major regression is a big deal. Missing a tiny regression is less important, and may be overall better than spending a lot of time investigating false alarms.}{P5}

In contrast, minimizing false alerts yields more divided opinions (\textit{Q10}), with participants equally split on agreeing and disagreeing with the premise of increasing the technique's precision.
This tension reflects the inherent difficulty of the trade-off and helps explain the seemingly neutral stance in \textit{Q9} alongside the stronger preference patterns observed in \textit{Q10} and \textit{Q11}.
Nevertheless, the cost of false alerts is not dismissed entirely. 
\practitionerquote{False alerts come with significant cost and reduce confidence in the system, which is problematic.}{P1}

\paragraph{Assessment of the \currentmozillamethod performance.}
The higher focus on minimizing false negatives is also reflected in the practitioners' assessment of \currentmozillamethod. 
Respondents are largely satisfied with its ability to minimize false alerts (\textit{Q13}), with ten out of twelve agreeing or strongly agreeing, a sentiment well-grounded in \currentmozillamethod's 99.2\% precision reported in Section \ref{sec:results}. 
However, \currentmozillamethod's handling of missing alerts (\textit{Q14}) is rejected as nine out of twelve respondents disagree or strongly disagree with being acceptable in practice.
Multiple practitioners point directly to the $\sim$42\% miss rate implied by this recall figure. 
\practitionerquote{Current method has a number of missed alerts that almost equals the number of correct alerts. If there are regressions among those missed alerts, it's a very high number.}{P7}
\practitionerquote{The current method is great at avoiding false alerts, but that comes at the cost of missing real regressions. A 42\% change slipping through undetected is a clear sign the system is too conservative.}{P11}
The practitioner consensus, acknowledging \currentmozillamethod's near-perfect precision while condemning its recall, provides strong empirical motivation for the work presented in this paper.

\summarybox{Validating Research Assumptions}{
Practitioners strongly lean toward minimizing missed alerts (11 out of 12 endorse this), driven by direct user impact, though the cost of false alerts is acknowledged, as it erodes engineering confidence.
Finally, practitioners approve of \currentmozillamethod's near-perfect precision (99.2\%) but strongly agree that the method's recall (57.7\%) should be improved. }

\subsection{Validating CPD methods Results}
\label{subsec:validating_cpd_results}

\begin{figure}
    \centering
    \includegraphics[width=0.9\columnwidth]{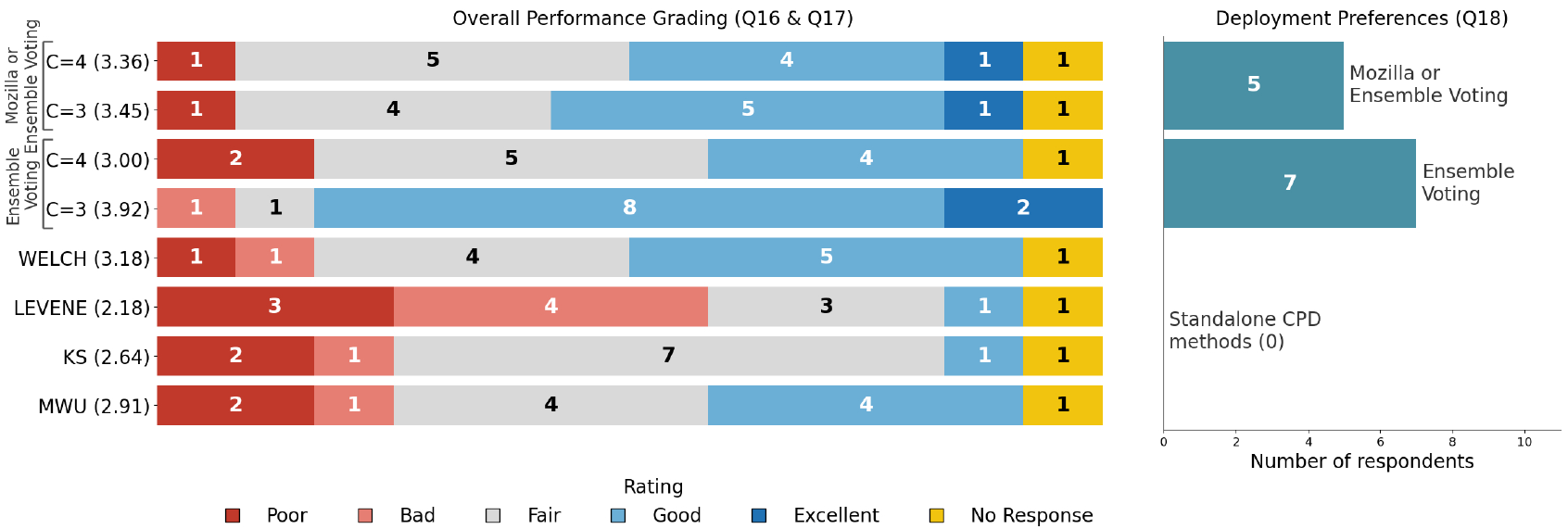}
    \caption{MOS Performance Grading (Q16 \& Q17) and Deployment Preferences (Q18)}
    \label{fig:MOS_test_results}
\end{figure}

\noindent
\textbf{Ratings of Proposed Methods.}
The left subfigure of Figure~\ref{fig:MOS_test_results} reports the MOS ratings collected in \textit{Q16} and \textit{Q17} for the standalone statistical methods and voting strategies, respectively.
The voting strategies (\textit{Q17}) are rated considerably more favorably than standalone methods. 
\textit{Ensemble voting (C=3)} receives the highest mean score across all evaluated techniques, with 8 respondents rating it \textit{Good} and 2 rating it \textit{Excellent}, consistent with its strong recall improvement over the baseline. 
\textit{Mozilla or Ensemble voting (C=3)} and \textit{Mozilla or Ensemble voting (C=4)} follow closely, both receiving predominantly \textit{Good} ratings, reflecting appreciation for strategies that recover missed detections while preserving much of \currentmozillamethod's precision.                                                            
\textit{Ensemble voting (C=4)} received a relatively lower rating, a profile that practitioners appear to find overly cautious, given the operational priority they place on minimizing missed regressions.

Among the standalone methods, ratings vary more than the voting strategies.
\texttt{WELCH} earns a mean MOS score of 3.18, comparable to \textit{Ensemble voting (C=4)}.
However, responses concentrate around \textit{Fair} and \textit{Good} with no \textit{Excellent} ratings, and its best-configuration F1-score of 67.1\% still falls short of Mozilla's baseline of 70.6\%.
\texttt{KS} and \texttt{MWU} are rated predominantly \textit{Fair}, reflecting limited confidence in methods that improve recall without a proportional gain in overall F1-score.
\texttt{LEVENE} receives the lowest ratings (2.18 out of 5) among all evaluated techniques, consistent with its F1-score of only 49.3\% and recall of 51.7\%, making it the weakest individual method on both dimensions.
\\

\noindent
\textbf{Deployment Preference.}
Practitioners were asked which group of techniques they consider adequate for deployment on \textit{Perfherder} as shown in the right subfigure of Figure \ref{fig:MOS_test_results}. 
Respondents clearly converge on voting strategies: seven select \textit{Ensemble voting} and five select \textit{Mozilla or Ensemble voting}, while no respondent endorses any standalone CPD method. Two respondents do not provide any response about their preference. 
The free-text responses illustrate some reasoning behind these preferences, as practitioners emphasize the importance of a technique with a better recall:

\practitionerquote{I believe Mozilla’s top priority should be detecting as many regression alerts as possible. This means minimizing missed alerts, even if it results in more false positives. Sheriffs are responsible for evaluating these alerts, confirming whether they are valid regressions or marking them as invalid. Although this may increase their workload, it ultimately leads to better performance quality and a more reliable product.}{P10}

Indeed, relative to \currentmozillamethod, the ensemble strategy gains 27.9 percentage points in recall at the cost of a 13.8\% reduction in precision, a trade-off that most practitioners consider acceptable given the operational priority placed on minimizing missed regressions.

\practitionerquote{I favor systems which lead to higher correct alerts. For dealing with the false positives, we can improve our tooling, technology, and process to minimize the cost.}{P2}

\summarybox{Practitioner's perspective on CPD results in practice.}{
Voting strategies receive substantially higher practitioner ratings and endorsement for deployment than standalone methods, with \textit{Ensemble voting (C=3)} earning the strongest endorsement across all evaluated techniques.       
Practitioners accept the precision trade-off that voting strategies introduce, viewing the significant gains in recall as essential given the operational cost of missed regressions.
}

\section{\textbf{Discussion}}
\label{sec:discussion}

We discuss the findings of our study across three subsections. 
We first discuss the practical implications of our benchmarking results for CPD method selection and tuning, then examine how practitioners perceive the proposed improvements, and finally reflect on the limitations to consider when interpreting our conclusions.

\subsection{Implications for Researchers}
\label{subsec:implications_for_researchers}

Our results highlight several methodological considerations for researchers working on CPD benchmarking and evaluation. 

\noindent
\textbf{The Importance of Method Calibration.} First, the large gap between the \textit{Default} and \textit{Upper-Bound Performance} configurations observed across all method families reveals that many methods underperform, not due to fundamental limitations but due to poor default calibration. This is consistent with the findings of ~\citet{TCPDBenchAlanTuringInstitute}, who report similar gaps between default and oracle configurations in their benchmark and note that methods such as \texttt{kcpa} and \texttt{rfpop} only become competitive after tuning. These results imply that benchmarking CPD methods under fixed default parameters risks systematically misrepresenting their true capabilities, and that investing in parameter selection is critical for obtaining the best performance of these CPDs.

\noindent
\textbf{Accounting for inter-annotator disagreement in CPD benchmarks.} A broader implication of our annotation study is that collapsing multi-annotator labels into a single ground truth discards a meaningful signal. The disagreement we observe is not random noise but reflects genuine ambiguity in performance data, as evidenced by its correlation with perceived difficulty as stated in Section \ref{subsec:annotation_data_quality_assessment}. In our study, we account for this by adopting a multi-annotator evaluation framework following~\citet{Martin2004}. Future CPD benchmark construction efforts should similarly model and report inter-annotator disagreement explicitly.
\\

\noindent
\textbf{Survey design and response consistency}
Practitioners differ on whether to prioritize precision or recall in change point detection methods. When asked abstractly in \textit{Q9} whether false and missing alerts deserve equal weight, respondents were evenly split. Yet when asked independently in \textit{Q10} and \textit{Q11} whether each error type should be prioritized, eleven out of twelve strongly endorsed minimizing missed alerts, suggesting a general lean toward recall in practice. One possible explanation is that respondents had already been exposed to \currentmozillamethod's performance profile, specifically its 99.2\% precision and 57.7\% recall prior to filling the survey, which may have anchored their responses toward the known weakness of the system. Regardless, this divergence highlights that practitioner priorities are not uniform. Future studies evaluating change point detection methods should therefore report not only the standard F1 score but also recall-weighted variants such as the F2 score, alongside precision-weighted alternatives, to ensure results can be meaningfully interpreted across different operational contexts and priority profiles.

\subsection{Implications for Practitioners}
\label{subsec:implications_for_practitioners}

Our findings carry several actionable implications for practitioners responsible for deploying and maintaining performance alerting systems. 

\noindent
\textbf{Ensemble methods are a good starting point.}
The practitioner survey showed that no standalone CPD method was endorsed for deployment, whereas ensemble voting strategies were deemed as a good compromise between precision and recall. 
This consensus reflects the quantitative results, where the offline ensemble peaked at 78.4\% F1-score, and \currentmozillamethod or Ensemble strategy reached 77.2\%, both substantially outperforming any individual method. Practitioners should therefore treat ensemble voting not as an experimental extension but as the primary deployment candidate when upgrading existing CPD pipelines. For offline ensembles, we suggest starting with \texttt{BinSeg}, \texttt{CPNP}, \texttt{KCPA}, and \texttt{MongoDB} as the initial set of methods to experiment with, as this combination achieved the best F1-score in our evaluation. For hybrid ensembles, we recommend experimenting with all available methods except \texttt{LEVENE}, which yields the worst F1-score, precision, and recall among all hybrid methods evaluated and is therefore unlikely to contribute meaningfully to the ensemble's detection quality.

\noindent
\textbf{Explicitly quantify the precision-recall tradeoff.}
Beyond method selection, our results suggest that the precision-recall tradeoff should be treated as an explicit organizational policy rather than an implicit consequence of threshold defaults. 
While survey respondents were evenly split when asked abstractly whether both error types deserve equal weight, eleven out of twelve strongly endorsed minimizing missed alerts when asked independently, driven by the direct user impact of undetected regressions. 
Organizations should explicitly encode this priority in their alerting configurations rather than relying on default parameter choices. 
\\

\noindent
\textbf{The hidden issue of false negatives.} 
Finally, practitioners should be cautious about interpreting low manual alert creation rates as evidence that their automated system is performing well. In our preliminary analysis, only 6.8\% of alert summaries contained manually created alerts suggesting missed detections, a figure that may appear reassuring in isolation. However, our evaluation of Mozilla's method reveals a recall of only 57.7\%, meaning that roughly four in ten annotated regressions go undetected. This discrepancy is consistent with prior work showing that practitioners tend to under-report automation failures when they place high trust in the automated system~\cite{fn_motivation_1, trust_in_seq_non_param_tests, trust_in_auto_study}. The low rate of human-detected misses therefore likely understates the true prevalence of undetected regressions, and practitioners should not rely on manual creation rates alone as a proxy for detection coverage.

\subsection{Lessons Learned From the Integration and Deployment}
\label{subsec:deployment_lessons}

The integration began after consolidating the results of the practitioner survey, and remains an ongoing process.
In the process, we face a series of challenges that require changes in the original chosen method. 
We discuss these challenges in the remainder of the section.

\textbf{Constrain the methods to the same window size.} 
For better explainability, we opt to deploy a voting system in which all methods use the same data window size, so that all individual methods analyze the same data. This design choice was motivated by practitioners contributing to the voting system deployment, who emphasize that when all methods operate on the same window, the p-values they produce are directly comparable, ensuring that the ensemble consolidates conclusions drawn from the same observations, rather than mixing results across different data ranges. It is worth noting that the results reported in this paper and presented in the practitioner survey were obtained without this constraint, meaning that the deployed configuration differs from the evaluated one in this regard.

\textbf{Restricting tolerance margin further.}
Although most respondents did not object to a tolerance margin of M=5 revisions, it proved too large in practice. Specifically, a senior engineer involved in the voting system integration pointed out that a margin of 5 corresponds to a full day's worth of performance measurements, which represents too wide a window given that multiple regressions can be introduced within a single day. We therefore adopted M=1 instead based on that feedback, as a wider window risks merging two closely occurring regressions into a single alert, making it harder for teams to identify and address each issue separately.

\textbf{Levene instability}. For hybrid ensembles, we recommend experimenting with all available methods except \texttt{LEVENE}. Beyond yielding the weakest F1-score, precision, and recall among all hybrid methods evaluated, it exhibited implementation instability during integration testing, producing unit test failures across multiple configurations. These issues further motivated its exclusion from the deployed ensemble.

\textbf{Recurring adjacent alerts.} During integration, several methods began generating alerts from adjacent revisions. Upon investigation, the root cause was traced to p-values of exactly zero returned by multiple methods for consecutive revision pairs. This behavior stems from a subtle asymmetry between T-value and p-value comparison logic. In the original system, when two adjacent revisions both exceed the T-value threshold, the one with the higher T-value is selected. When switching to p-value-based comparisons, the logic is inverted: a change point is flagged when the p-value falls \textit{below} a threshold rather than above it. Since p-values are strictly positive by definition, a returned value of exactly zero indicates numerical underflow: the computed p-value is so small that it falls below the floating-point representation limit. This caused multiple adjacent revisions to be simultaneously flagged, as the deduplication logic originally designed for T-values did not account for this edge case. Notably, this issue did occur in the experimental setup on some occasions but went undetected prior to deployment. It was addressed by retaining only the last revision among those yielding a p-value of zero, effectively applying a conservative deduplication strategy for underflow cases.

\section{Threats To Validity}
\label{sec:threats_to_validity}

We discuss the main threats to the validity of our findings, organized along three dimensions: construct, internal, and external validity.
Being mindful of recent criticism of the appropriateness of the threats to validity framing for software engineering research~\cite{verdecchia2023threats}, as part of the discussion below we reflect on trade-offs inherent in designing empirical studies as recommended by Robillard et al~\cite{robillard2024communicating}. 

\subsection{Construct Validity}

Several choices in our evaluation design may affect how accurately our metrics reflect real-world detection quality. 

\noindent
\textbf{Evaluation metric.} 
First, the F1-score treats false positives and false negatives as equally costly, an assumption that \textit{Q9} does not explicitly reject, though the asymmetry observed in \textit{Q10} and \textit{Q11} suggests practitioners may operationally prioritize recall. This asymmetry may itself reflect practitioners reasoning from familiarity with the current system's known weaknesses rather than a principled stance against equal weighting, yet the possibility remains that a recall-weighted variant such as an F-beta score could yield different method rankings. 

\noindent
\textbf{Tolerance margin.}
We define a detection as correct if it falls within five revisions of a ground-truth annotation. While this choice introduces some ambiguity into what constitutes a true positive, and a stricter margin could penalize methods differently, we partially mitigate this threat by validating the margin with practitioners, seven out of twelve of whom endorsed M=5 as a pragmatic operational tolerance. Furthermore, as discussed in Section~\ref{sec:discussion}, we additionally adopted M=1 during deployment integration following practitioner feedback, providing a complementary perspective on method performance under a stricter criterion.

\noindent
\textbf{Subjective ground-truth.} Ground-truth annotation is inherently subjective: even among eleven experienced Mozilla engineers, disagreement on the exact location or existence of a change point is possible. We partially mitigate this threat by having each timeseries annotated by five distinct practitioners on average, ensuring that the ground truth reflects a collective judgment rather than a single individual's perspective. The level of agreement observed in the annotations further supports the reliability of this ground truth: out of 5781 annotated changes, 24.48\% were agreed upon by all annotators, and the mean OvR F1 score across all tasks is 0.76 with a median of 0.79 as stated in Section~\ref{subsec:annotation_data_quality_assessment}, indicating that annotators generally agree on the location of change points. Residual disagreement is most pronounced on harder timeseries, where perceived difficulty correlates negatively with agreement (Pearson r=-0.42), suggesting that subjectivity is concentrated in genuinely ambiguous cases rather than being a systematic bias across the dataset.

\subsection{Internal Validity}

\noindent
\textbf{CPD Methods Implementations}. Several factors internal to our experimental design may affect the reliability of our results. Regarding implementation fidelity, porting methods into the TCPDBench framework introduces a risk of compatibility or configuration errors. We mitigate this for Mozilla's method by validating our replication against 30 manually cleaned time series and confirming that it reproduces \textit{Perfherder}'s output exactly. We also perform sanity checks on third-party algorithms to detect obvious over- or under-reporting, yet subtle implementation discrepancies in less well-documented methods cannot be fully ruled out.

\noindent
\textbf{Limitations on the Hyper-parameter tuning.} 
Two related design choices limit the conclusions that can be drawn from the ensemble results. First, voting experiments are conducted using only the single best-performing configuration per method rather than searching the joint hyperparameter space across all ensemble members, meaning the true optimal ensemble configuration remains unexplored. Second, the hyperparameter search spaces differ substantially across methods, ranging from as few as four combinations for \texttt{RFPOP} to 792 for hybrid methods, which means methods with larger grids have a structurally higher chance of identifying a well-performing configuration. This disparity makes direct cross-method comparisons under tuned settings potentially inequitable, and some methods evaluated on restricted grids may have achieved higher F1-scores had a more exhaustive search been conducted.

\subsection{External Validity}

Several factors limit the generalizability of our findings beyond the specific context in which this study was conducted. 
The study is deeply embedded in Mozilla's CI/CD workflow and tooling, and the relative performance of CPD methods may differ in organizations with different code change frequencies, testing cadences, or noise profiles. Within Mozilla itself, the ground-truth dataset of 174 time series is drawn exclusively from two test subsets, Speedometer3 and TP6, selected for their reliability, meaning that detection performance on more volatile or heterogeneous Mozilla test suites may diverge from what we report. 
The practitioner validation is subject to a related limitation as it relies on a single workshop and follow-up survey, which, while providing valuable industrial grounding, falls short of the depth achievable through more rigorous elicitation methods such as multi-round Delphi studies or individual interviews, and its conclusions should therefore be interpreted as indicative rather than definitive.

\section{\textbf{Related Work}}
\label{sec:related_works}

This section reviews prior work on software performance engineering, focusing on the anomaly detection aspect. These works provide context for our contribution.

\subsection{CPD Methods in Software Performance}
\label{subsec:perf_anomaly_detetion_methods_related_works}

A substantial body of research has examined the usage of statistical CPD methods for anomaly detection in software performance data.
\citet{mongodb_cpd} reports that MongoDB relies on detecting performance changes through a statistical method called E-divisive meanings~\cite{ecp}, where the system generates a type of alert once a change is detected based on threshold comparisons. This implementation reduces the number of false positive alerts from 99\% to 40-80\%. The same statistical method is used by~\citet{ingo20258yearsoptimizingapache} to identify performance regressions in benchmark data.

CUSUM (Cumulative Sum Control Chart) method is employed by~\citet{muehlbauer2019accurate} to identify change points in the performance evolution of six real-world software systems. The analysis focuses on detecting abrupt and substantial performance changes by computing the cumulative deviation from a target value across revisions. Another work by ~\citet{lee2012} explores the use of CUSUM and Shewhart control charts to identify performance anomalies in database systems during continuous integration. The approach relies on detecting both large and subtle performance shifts introduced by code changes. It is found that CUSUM can reliably detect small performance changes while maintaining a low false alarm rate comparable to Shewhart. We adopt these methods as part of the online methods in our study.

\citet{cloudplatformapp} proposes a work focusing on the performance of Cloud platform applications, the efficiency of statistical CPD methods such as PELT and Binary Segmentation are evaluated within a broader anomaly detection and root cause analysis framework. Overall, the methods effectively detect sustained workload changes and help attribute anomalies to root causes, demonstrating their efficiency in real-world performance monitoring.

\citet{little_law_approach} propose a technique for proactively detecting performance anomalies by leveraging the inverse relationship between throughput and response time established by Little’s Law. Their approach utilizes Individuals and Moving Range (XmR) control charts to automatically flag early-warning signals when throughput drops below a lower control limit while response times simultaneously exceed an upper limit.

Our work consolidates such CPD methods that follow several statistical methods' families such as offline and online detection into the established and openly available \textit{TCPDBench} toolset for anyone to utilize for software performance data specifically.

Complementing these statistical foundations, Machine Learning-based change point detection is also adopted. \citet{testleveljustintime} presents an approach that automatically predicts whether a specific code commit will cause a test to manifest performance regressions is developed. It builds random forest classifiers which combine multiple tree predictors, trained on all prior commits to identify potential regressions, achieving high predictive accuracy with an average AUC of 0.86. Complementing this, \textit{AutoPerf} is introduced by~\citet{alam2020zeropositivelearningapproachdiagnosing}. It is a tool designed to automate regression testing which relies on autoencoder neural networks and k-means clustering for change point detection in software performance data, allowing the tool to accurately diagnose more performance bugs than previous state-of-the-art approaches.
Despite the high predictive accuracy reported by these ML-based systems, our study focuses exclusively on statistical CPD methods for three primary reasons. First, statistical methods do not require the large-scale labeled training sets that ML models necessitate, making them more adaptable to new projects. Second, they offer superior transparency and interpretability, allowing engineers to understand exactly why a regression was flagged.

\subsection{Practitioner Validation in Software Engineering Research}
\label{subsec:practitioner_alidation_in_software_engineering_research}

\citet{forepost_paper} presents \textit{FOREPOST}, an automated, black-box testing system designed to identify performance bottlenecks by learning from application execution traces. Its effectiveness is validated by industry experts who manually confirm its top 30 identified bottlenecks as genuine. Notably, the system uncovers a resource-intensive bug in a Visitor pattern implementation that, when fixed, results in a 7\% improvement in overall application performance.

\citet{prac_accepted_fixes} introduces \textit{Diff-Design Structure Matrix}, a modeling technique evaluated through an empirical study of 192 performance issues from five Apache projects, using solutions already accepted by industry practitioners. To ensure data integrity, two researchers independently verify the issues to exclude false positives and focus strictly on intentional performance optimizations. This empirical grounding validates the system's ability to model the complex structural changes practitioners use to resolve architectural bottlenecks. By analyzing the Return on Investment of these expert-approved fixes, the study provides a framework for practitioners to prioritize optimization strategies based on measurable effort and performance gains.

\citet{little_law_approach} evaluate the effectiveness of their change point detection method through a 23-day production deployment on a large-scale, mission-critical case management system. During this evaluation period, the system successfully identifies all four documented slowness incidents, achieving 100\% recall and 100\% precision when configured to require two consecutive intervals of violation to filter out transient noise.

While prior industrial validations confirm detected issues through expert review~\cite{forepost_paper}, practitioner-accepted fixes~\cite{prac_accepted_fixes}, or production deployment~\cite{little_law_approach}, none explicitly involve practitioners in evaluating the research assumptions underpinning the evaluation framework itself. Our work goes further: we consult Mozilla performance practitioners not only to assess the deployment readiness of our proposed methods, but also to validate research assumptions such as the tolerance margin and error-weighting assumptions central to our evaluation design which is an aspect largely absent from existing industrial validation efforts in performance engineering.

\section{\textbf{Conclusions}}
\label{sec:conclusion}

In this study, we first record the baseline performance of \currentmozillamethod for performance regressions detection, which we then use as a benchmark for comparison. We proceed with an extensive annotation collection phase, involving performance experts to establish a ground-truth dataset of 174 time series. This dataset enables rigorous evaluation of CPD methods, in which we analyze the trade-offs among 25 statistical algorithms and 15 ensemble strategies. Finally, we perform a validation with practitioners, whose feedback confirmed the practical utility of our findings.

Our results demonstrate that while traditional statistical methods provide a foundation for regression detection, their effectiveness in a production environment depends on minimizing false positives without sacrificing sensitivity. We found that individual algorithms often struggle to meet these requirements. However, ensemble voting strategies, particularly those that leverage the existing high-precision baseline, achieve a significant performance increase, reaching an F1-score of 78.4\%.
Practitioner validation further confirms that ensemble voting strategies are the preferred deployment candidates, with all twelve respondents endorsing voting-based approaches over standalone methods and none selecting any individual CPD method as deployment-ready. 
The dataset, toolset extensions, and validated ensemble strategies produced by this study offer a reusable foundation for researchers and practitioners working to improve the reliability of change point detection in real-world performance monitoring pipelines.

\section{\textbf{Acknowledgments}}
\label{sec:acknowledgments}
We acknowledge the support of \textit{Mitacs} and the \textit{Mitacs Accelerate} Program. We also acknowledge the support of the Natural Sciences and Engineering Research Council of Canada (NSERC). The authors thank Michele Tucci from University of L'Aquila for his guidance.

\bibliographystyle{ACM-Reference-Format}
\bibliography{bibliography_pl}

\appendix

\end{document}